\newcommand{\manuscript}{3}

\newcommand{\psSurveyArea}{600}
\newcommand{\psSurveyAreaEqu}{300}
\newcommand{\psSurveyAreaSouth}{292}

\newcommand{\cl}{C_\ell}
\newcommand{\nhat}{\hat{\mathbf{n}}}

\newcommand{\ave}[1]{\left\langle#1\right\rangle}

\newcommand{\LCDM}{$\Lambda$CDM }

\newcommand{\be}{\begin{equation}}
\newcommand{\ee}{\end{equation}}
\newcommand{\ba}{\begin{eqnarray}}
\newcommand{\ea}{\end{eqnarray}}
\newcommand{\ara}{148}
\newcommand{\arb}{218}

\newcommand{\actE}{ACT-E}
\newcommand{\actS}{ACT-S}

\newcommand{\avg}[1]{\left< #1 \right>} 
\newcommand{\nobs}{\ensuremath{n_{\mathrm{obs}}}}
\newcommand{\lensSigma}{4.6}
\ifnum\manuscript=2
\documentclass[11pt, letter]{article}

\newcommand{\shorttitle}[1]{}
\newcommand{\shortauthors}[1]{}
\newcommand{\received}[1]{}
\newcommand{\altaffiltext}[2]{#1 #2}
\newcommand{\altaffilmark}[1]{\ensuremath{^{\mathrm{#1}}}}
\newcommand{\keywords}[1]{\emph{keywords}---#1}
\newcommand{\acknowledgements}[1]{ }
\renewcommand{\author}[1]{#1}
\renewcommand{\title}[1]{{\centering {\Large \bf #1}\\ }}

\else\ifnum\manuscript=1
\documentclass[12pt,letterpaper,preprint,flushrt]{aastex}
\else\ifnum\manuscript=3
\documentclass{emulateapj}
\usepackage[citebordercolor={0 .5 .5}]{hyperref}

\else
\documentclass{emulateapj}
\fi
\usepackage{graphicx}        % AASTeX preferred
\usepackage{bm}              % boldface math
\usepackage{natbib}
\usepackage{mathrsfs}
\newcommand{\arone}{148\,GHz}
\newcommand{\artwo}{218\,GHz}
\newcommand{\arthree}{277\,GHz}

\newcommand{\commentx}[1]{}

\renewcommand{\vec}[1]{\mbox{\boldmath$#1$}} % Bold
\newcommand{\etal}{et al.\,}  % Dude: it's not italicized.  See 15th Chicago
\newcommand{\ra}[3]   % right ascension
   {\makebox[1.5em][r]{#1}\makebox[1.5em][r]{#2} \makebox[2em][r]{#3}}

\newcommand{\hms}[3]  % Write HMS in ApJ style.
   {${#1}^{\mathrm{h}}{#2}^{\mathrm{m}}{#3}^{\mathrm{s}}$}

\newcommand{\hmin}[2]  % Write HM in ApJ style.
   {\ensuremath{{#1}^{\mathrm{h}}{#2}^{\mathrm{m}}}}
   
\newcommand{\hours}[1]  % Write H in ApJ style.
   {\ensuremath{{#1}^{\mathrm{h}}}}

\newcommand{\dms}[3]  % Write DMS in ApJ style.
   {\ensuremath{{#1}\degree{#2}\arcminute{#3}\arcsecond}}

\newcommand{\dm}[2]  % Write DM in ApJ style.
   {\ensuremath{{#1}\degree{#2}\arcminute}}

\newcommand{\ukcmb}  % microKelvin_cmb
           {\ensuremath{\micro \kelvin_\mathrm{cmb}}}

\newcommand{\uk}  % microKelvin
           {\ensuremath{\micro \kelvin}}

\newcommand{\fdeg} % fractional degrees
           {\hbox{$.\!\!^{\circ}$}}

\usepackage{color}

\usepackage[mediumspace,Gray,squaren]{SIunits}
\usepackage{rotating}
\usepackage{graphicx}
\usepackage{amsmath}
\shorttitle{ACT Three-season Power Spectrum}
\shortauthors{S. Das \etal}

\begin{document}
\title{The Atacama Cosmology Telescope: Temperature and Gravitational Lensing Power Spectrum Measurements  
from Three Seasons of  Data}
%%%%%%%%%%%%%%%%%%%%%%%%%%%%%%%%%%%%%%%%%%%%%%%%%%%%%%%%%%%%%%%%%%
%%%%%%%%%
% WARNING: This LaTeX block was generated automatically by authors.py
% Do not change by hand: your changes will be lost.

%%%%%%%%%%%%%%%%%%%%%%%%%%%%%%%%%%%%%%%%%%%%%%%%%%%%%%%%%%%%%%%%%%%%%%%%%%%
% WARNING: This LaTeX block was generated automatically by authors_das.py
% Do not change by hand: your changes will be lost.

\author{
Sudeep~Das\altaffilmark{1,2},
Thibaut~Louis\altaffilmark{3},
Michael~R.~Nolta\altaffilmark{4},
Graeme~E.~Addison\altaffilmark{5,3},
Elia~S.~Battistelli\altaffilmark{6,5},
J~Richard~Bond\altaffilmark{4},
Erminia~Calabrese\altaffilmark{3},
Devin~Crichton\altaffilmark{7},
Mark~J.~Devlin\altaffilmark{8},
Simon~Dicker\altaffilmark{8},
Joanna~Dunkley\altaffilmark{3},
Rolando~D\"{u}nner\altaffilmark{9},
Joseph~W.~Fowler\altaffilmark{10,11},
Megan~Gralla\altaffilmark{7},
Amir~Hajian\altaffilmark{4},
Mark~Halpern\altaffilmark{5},
Matthew~Hasselfield\altaffilmark{12,5},
Matt~Hilton\altaffilmark{13,14},
Adam~D.~Hincks\altaffilmark{4},
Ren\'ee~Hlozek\altaffilmark{12},
Kevin~M.~Huffenberger\altaffilmark{15},
John~P.~Hughes\altaffilmark{16},
Kent~D.~Irwin\altaffilmark{10},
Arthur~Kosowsky\altaffilmark{17},
Robert~H.~Lupton\altaffilmark{12},
Tobias~A.~Marriage\altaffilmark{7,12},
Danica~Marsden\altaffilmark{18,8},
Felipe~Menanteau\altaffilmark{16},
Kavilan~Moodley\altaffilmark{14},
Michael~D.~Niemack\altaffilmark{19,10,11},
Lyman~A.~Page\altaffilmark{11},
Bruce~Partridge\altaffilmark{20},
Erik~D.~Reese\altaffilmark{8},
Benjamin~L.~Schmitt\altaffilmark{8},
Neelima~Sehgal\altaffilmark{21},
Blake~D.~Sherwin\altaffilmark{11},
Jonathan~L.~Sievers\altaffilmark{11,4},
David~N.~Spergel\altaffilmark{12},
Suzanne~T.~Staggs\altaffilmark{11},
Daniel~S.~Swetz\altaffilmark{10},
Eric~R.~Switzer\altaffilmark{4},
Robert~Thornton\altaffilmark{22,8},
Hy~Trac\altaffilmark{23},
Ed~Wollack\altaffilmark{24}
}
\altaffiltext{1}{Argonne National Laboratory, 9700 S.~Cass Ave., Lemont, IL 60439}
\altaffiltext{2}{Berkeley Center for Cosmological Physics, LBL and
Department of Physics, University of California, Berkeley, CA, USA 94720}
\altaffiltext{3}{Sub-department of Astrophysics, University of Oxford, Keble Road, Oxford, OX1 3RH, UK}
\altaffiltext{4}{Canadian Institute for Theoretical Astrophysics, University of
Toronto, Toronto, ON, Canada M5S 3H8}
\altaffiltext{5}{Department of Physics and Astronomy, University of
British Columbia, Vancouver, BC, Canada V6T 1Z4}
\altaffiltext{6}{Department of Physics, University of Rome ``La Sapienza'', 
Piazzale Aldo Moro 5, I-00185 Rome, Italy}
\altaffiltext{7}{Dept. of Physics and Astronomy, The Johns Hopkins University, 3400 N. Charles St., Baltimore, MD 21218-2686}
\altaffiltext{8}{Department of Physics and Astronomy, University of
Pennsylvania, 209 South 33rd Street, Philadelphia, PA, USA 19104}
\altaffiltext{9}{Departamento de Astronom{\'{i}}a y Astrof{\'{i}}sica, 
Facultad de F{\'{i}}sica, Pontific\'{i}a Universidad Cat\'{o}lica,
Casilla 306, Santiago 22, Chile}
\altaffiltext{10}{NIST Quantum Devices Group, 325
Broadway Mailcode 817.03, Boulder, CO, USA 80305}
\altaffiltext{11}{Joseph Henry Laboratories of Physics, Jadwin Hall,
Princeton University, Princeton, NJ, USA 08544}
\altaffiltext{12}{Department of Astrophysical Sciences, Peyton Hall, 
Princeton University, Princeton, NJ USA 08544}
\altaffiltext{13}{Centre for Astronomy \& Particle Theory, School of Physics and Astronomy, University of Nottingham, NG7 2RD, UK}
\altaffiltext{14}{Astrophysics \& Cosmology Research Unit, School of Mathematics, Statistics \& Computer Science, University of KwaZulu-Natal, Durban 4041, SA}
\altaffiltext{15}{Department of Physics, University of Miami, Coral Gables, 
FL, USA 33124}
\altaffiltext{16}{Department of Physics and Astronomy, Rutgers, 
The State University of New Jersey, Piscataway, NJ USA 08854-8019}
\altaffiltext{17}{Department of Physics and Astronomy, University of Pittsburgh, 
Pittsburgh, PA, USA 15260}
\altaffiltext{18}{University of California, Santa Barbara, CA, USA 93106}
\altaffiltext{19}{Department of Physics, Cornell University, Ithaca, NY, USA 14853}
\altaffiltext{20}{Department of Physics and Astronomy, Haverford College,
Haverford, PA, USA 19041}
\altaffiltext{21}{Stony Brook University. Physics and Astronomy Department, Stony Brook, NY 11794}
\altaffiltext{22}{Department of Physics, West Chester University of Pennsylvania, West Chester, PA, USA 19383}
\altaffiltext{23}{McWilliams Center for Cosmology, Wean Hall, Carnegie Mellon University, 5000 Forbes Ave., Pittsburgh PA 15213, USA}
\altaffiltext{24}{NASA/Goddard Space Flight Center,
Greenbelt, MD, USA 20771}

\begin{abstract}
We present the temperature power spectra of the cosmic microwave background (CMB)
derived from the three seasons of data from the Atacama Cosmology Telescope (ACT)  at 148 GHz and 218 GHz,
as well as the cross-frequency spectrum between the two channels.  We detect  and correct for contamination due to the Galactic cirrus in our equatorial  maps. We present the results of a number of tests for possible systematic error and conclude that any effects are not significant compared to the statistical errors we quote. Where they overlap, we cross-correlate the ACT and the South Pole Telescope (SPT) maps and show they are consistent. 
 The measurements  of  higher-order peaks in the CMB power spectrum provide an additional test of  the \LCDM cosmological model, and 
help constrain extensions beyond the standard model.  The small angular scale power spectrum also provides constraining power 
on the Sunyaev-Zel'dovich effects and extragalactic foregrounds. We also present a  measurement of the CMB gravitational lensing convergence power spectrum  at  $\lensSigma\sigma$ detection  significance.
\end{abstract}


\setcounter{footnote}{0}

%%%%%%%%%%%%%%%%%%%%%%%%%%%%%%%%%%%%%%%%%%%%%%%%%%%%%%%
%%%%%%%%%
\section{INTRODUCTION}
% current status of PS measurements 
\defcitealias{das/etal:2011}{D11}
\defcitealias{das/etal:2011a}{DS11}
The current generation of  arcminute resolution cosmic microwave background (CMB)  experiments is 
providing researchers with a precise  view of CMB anisotropies over a range of scales ($500<\ell<10000$). Over the so-called \emph{Silk damping tail} of the CMB ($500<\ell<3000$) these observations
are revealing the subtle effects that inflationary physics,  primordial helium density and the energy density in relativistic 
degrees of freedom have on the acoustic oscillations in the photon-baryon plasma in the radiation-dominated era. Rapid progress in measurements of 
the  damping tail of the power spectrum has been achieved over a span  of a few years by a number of experiments, notably the Cosmic Background Imager (CBI; \citealt{sievers/etal:2009}),  Arcminute Cosmology Bolometer Array Receiver (ACBAR; \citealt{reichardt/etal:2009}), QUEST at DASI (QUaD; \citealt{brown/etal:2009, friedman/etal:2009}), the Atacama Cosmology Telescope (ACT; \citealt{swetz/etal:2010, fowler/etal:2010, das/etal:2011}, hereafter D11; \citealt{ dunkley/etal:2011}) and the South Pole Telescope (SPT; \citealt{ keisler/etal:2011, story/etal:prep}).
\begin{figure*}[t]
\begin{center}
\includegraphics[scale=0.34]{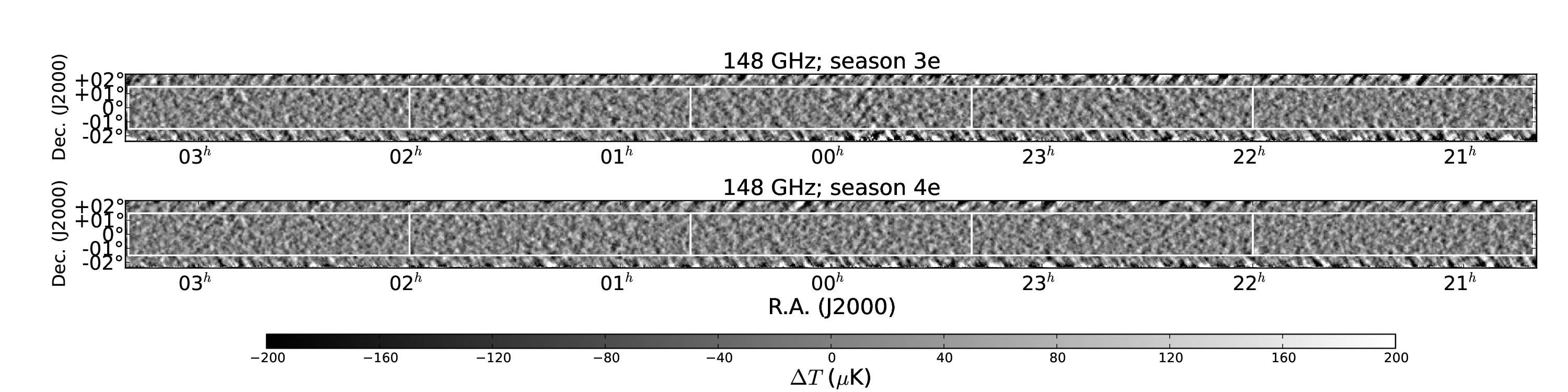}
\caption{Equatorial maps  (\actE) made from 2009 (upper panel) and 2010 (lower panel) \arone\ observations  filtered to emphasize modes in the range $\ell=500-2500$. The four data splits in either season were co-added to make this plot.  Also delineated are the patches used for computing power spectra.  }
\label{fig:equMaps}
\end{center}
\end{figure*}
 Minute  distortions of the CMB anisotropies  due to the gravitational lensing of CMB photons by large-scale structure have now been detected using
 CMB data alone, both in percent-level 
alteration of the damping tail acoustic peak structure (\citetalias{das/etal:2011}, \citealt{ keisler/etal:2011}, \citealt{story/etal:prep}), and in the subtle non-Gaussian signature  it induces in the statistics of CMB anisotropies \citep{das/etal:2011a, van/etal:2012}. When added to the cosmic variance-limited CMB power spectrum measurements at $\ell\lesssim 1000$ by the WMAP satellite \citep{larson/etal:2011}, the damping tail measurements are providing additional 
dynamic range resulting in improved constraints on inflationary parameters such as the tilt  and running of the primordial power spectrum.  On smaller scales ($\ell>3000$) the primordial CMB signal diminishes and emission from radio galaxies and  dusty star forming 
galaxies, as well as the thermal and kinematic Sunyaev-Zel'dovich effects \citep{sunyaev/zeldovich:1972} arising from the scattering of CMB photons by hot gas in galaxy clusters, dominate the power spectrum.  Along with the damping tail measurements, rapid progress has been made in the measurements
and modeling of this high-multipole tail (\citealt{lueker/etal:2010,shirokoff/etal:2011}; \citetalias{das/etal:2011}; \citealt{dunkley/etal:2011}, \citealt{reichardt/etal:2012}).  These measurements
have been used to estimate the thermal and kinematic SZ contributions to the power spectrum, as well as to model the cosmic infrared background (CIB) power spectrum arising from the mm-wave emission from dusty high redshift galaxies \citep{hall/etal:2010, dunkley/etal:2011, addison/etal:2012, 
addison/dunkley/bond:prep, hajian/etal:2012, reichardt/etal:2012, zahn/etal:2012}.\par
In this work,  we present the measurement of the power spectra of CMB temperature anisotropies at \arone{} and \artwo{} from a subset of 
ACT observations performed over the 2008, 2009 and 2010 observing seasons, and covering approximately  \psSurveyArea{}  deg$^2$ of the sky. 
This is approximately twice the survey area  used in a previous measurement of the ACT power spectrum\citepalias{das/etal:2011}.   Additionally, we also present an updated measurement of the gravitational lensing power spectrum from the equatorial strip. \citet{dunkley/etal:prep} use the temperature bandpowers reported in this work to generate  likelihood functions which  form the basis of the cosmological parameter constraints reported by Sievers et al. (2012).  \par

The paper  is organized as follows. In Section~\ref{sec:obs}  we describe the observations,  the generation of  maps from time 
ordered data, and the  estimation of the beam transfer functions.  We discuss the calibration of the maps in Section~\ref{sec:calib}. 
Section~\ref{sec:powspec} describes the pipeline used to process the maps into the  angular power spectrum. 
Treatment of point sources and other foregrounds is discussed in Section~\ref{sec:foreg}.   Simulations used to test  and validate various 
portions of the power spectrum estimation pipeline are described in Section~\ref{sec:sims}. The  power spectrum results and consistency checks are presented in Section~\ref{sec:results}, and the CMB lensing  results are discussed in  Section~\ref{sec:cmblensing}.    We conclude in Section~\ref{sec:conclusions}.

% Age of arcmin scale observations 
\defcitealias{fowler/etal:prep}{F10}
\defcitealias{Hincks/etal:prep}{H09}

\section{OBSERVATIONS AND FIELDS}
\label{sec:obs}
ACT is a  6-meter off-axis Gregorian telescope situated in the Atacama desert in  Chile at an elevation of 5190 m. ACT's  Millimeter 
Bolometric Array Camera (MBAC)  has three channels operating at \arone{},  \artwo{} and \arthree{}.  The instrument is described in detail in \citet{swetz/etal:2011}. 
 Between 2007 and 2010 ACT observed mainly along two constant-declination strips on the sky: one running along the celestial equator (hereafter the 
\emph{equatorial strip} or \actE), and the other along declination -55\degree\ in the southern sky (the \emph{southern 
strip} or \actS).    The observations were performed over  Nov 8 -- Dec 15, 2007;  Aug 11 -- Dec 24, 2008;  May 18 -- Dec 18, 2009, and  Apr 6 -- Dec 27, 2010. 
Here we present the angular power spectrum measurements of  \arone{} and \artwo{} observations from \psSurveyAreaEqu\ deg$^2$ on \actE\ and \psSurveyAreaSouth\ deg$^2$ on \actS. 
Table~\ref{table:obs} summarizes the observations from various seasons that were used in estimating the power spectrum, and also 
defines shorthand notations for maps, e.g., the 2009 season equatorial map is referred to as the ``season 3e'' or simply the ``3e'' map. 
\begin{deluxetable}{cccccc}
%\centering
\tablecolumns{6}
\tablewidth{0pc}
\tablecaption{Observations used in power spectrum estimation \label{table:obs}}
\tablehead{\colhead{Year} & \colhead{Key} &\colhead{ R.A. Range} & \colhead{Dec. Range}  & \colhead{Area} & \colhead{$n_p$\tablenotemark{a}}\\
\colhead{} & \colhead{} &  \colhead{} &  \colhead{} &\colhead{(deg$^2$)} & \colhead{}}
\startdata
\cutinhead{{Equatorial Strip \actE}}

%\hline
%\hline
2009 & 3e & \hmin{20}{40}  ..  \hmin{3}{20} & -1\fdeg5  ..  1\fdeg5  & 300 & 5\\
2010 & 4e & \hmin{20}{40}  ..  \hmin{3}{20} & -1\fdeg5  ..  1\fdeg5  & 300 & 5\\
%\line
%\hline 
\cutinhead{Southern Strip  \actS}
%\line
%\hline
2008f & 2sf &  \hmin{00}{22} .. \hmin{06}{47}& -55\fdeg0  ..  -50\fdeg0  & 292 & 4\\
2008 & 2s &  \hmin{04}{08} ..  \hmin{07}{08} & -55\fdeg2  ..  -51\fdeg2  & 146 & 2\\
2009 & 3s &  \hmin{04}{08} ..  \hmin{07}{08} & -55\fdeg2  ..  -51\fdeg2  & 146 & 2\\
2010 & 4s &  \hmin{04}{08} ..  \hmin{07}{08} & -55\fdeg2  ..  -51\fdeg2  & 146 & 2
\enddata
\tablenotetext{a}{number of patches}\\
\end{deluxetable}
%\end{table}
%%%%%%%%%%%%%%%%%

\subsection{Equatorial Observations}
\begin{figure}[htbp]
\begin{center}
\includegraphics[width=\columnwidth]{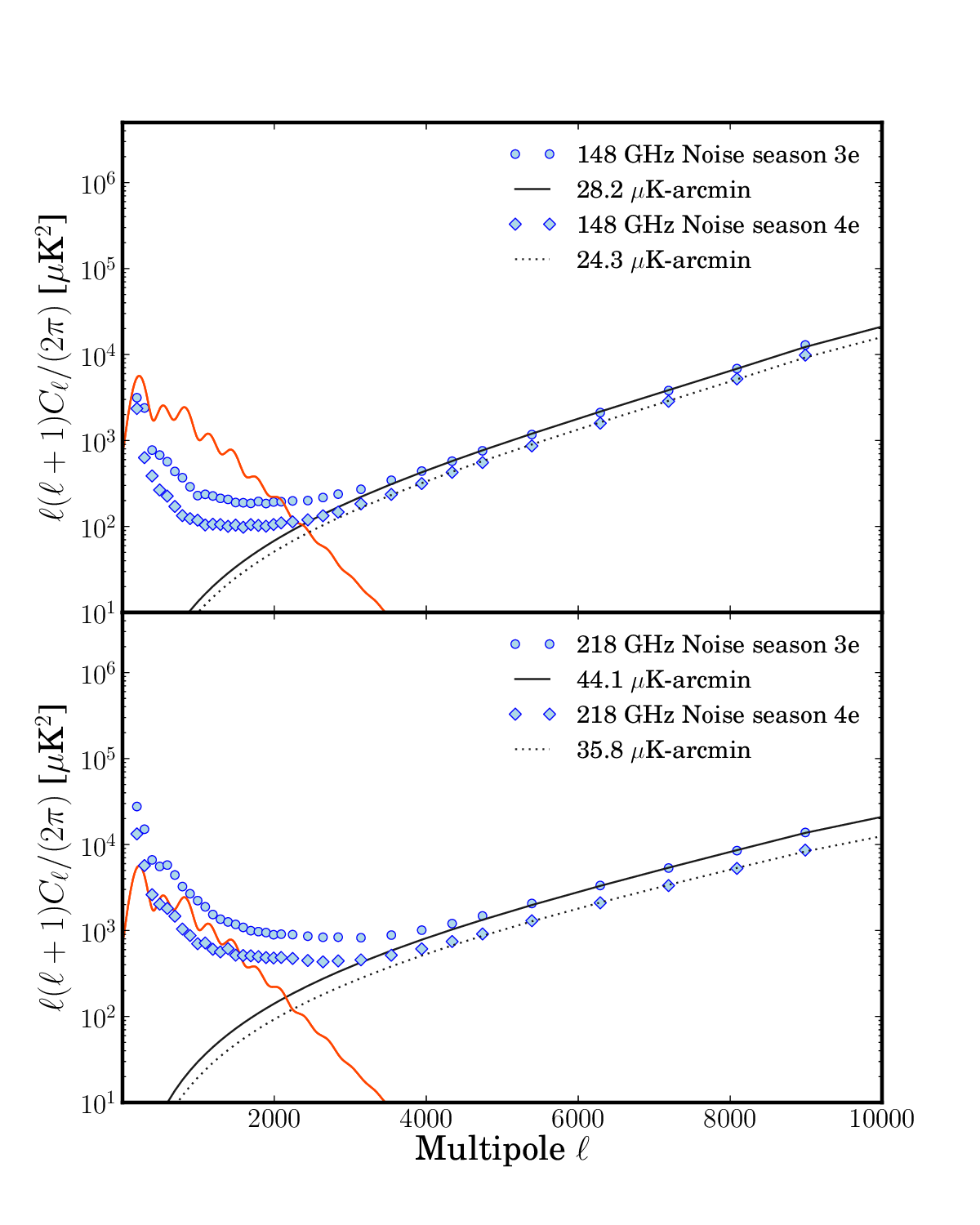}
\caption{Noise spectra for each season for the \actE\ maps  for \arone\ (upper panel) and \artwo\ (lower panel). 
The red solid line shows the CMB-only spectrum.  At \arone\ 
the power spectrum is sample variance limited at $\ell<2500$, while at \artwo\ detector and atmospheric noise dominate 
on most scales. \label{fig:equNoise}}
\end{center}
\end{figure}

Observations  on the \actE\ strip were performed in the 2009 and 2010 seasons, and  run along the celestial  equator with  a right ascension  span of  100 degrees, 
and a width of 3 degrees along the declination direction. For the power spectrum analysis, we make single season maps, and following \citetalias{das/etal:2011}   we divide the data within each season into four \emph{splits} in time,  by distributing data from roughly every fourth night into a different split,  generating four split-maps, each of which is properly cross-linked.  The maps are also spatially divided into five patches on which power spectrum estimation is performed separately. We explicitly avoid the edges of the maps where the cross-linking is poor and the noise is inhomogeneous.  A representation of the season 3e and season 4e \arone\ maps and patches are shown in Fig.~\ref{fig:equMaps}. The two seasons share the same footprint on the sky, and common 
patches were defined to facilitate the computation of cross season power spectra.  Fig.~\ref{fig:equNoise} shows the noise power spectra of the
\actE\ maps by season against the CMB-only theory.   For most seasons, and for \arone, on  largest angular scales ($\ell<500$) atmospheric noise dominates, 
while for intermediate angular scales ($500<\ell<2500$) fluctuations in the CMB dominate
 the variance.  At smaller angular  scales detector noise  becomes the most significant contribution. 
  
%%%%%%%%%%%%%%%
\subsection{Southern Observations}

\begin{figure*}[htbp]
\begin{center}
\includegraphics[scale=0.34]{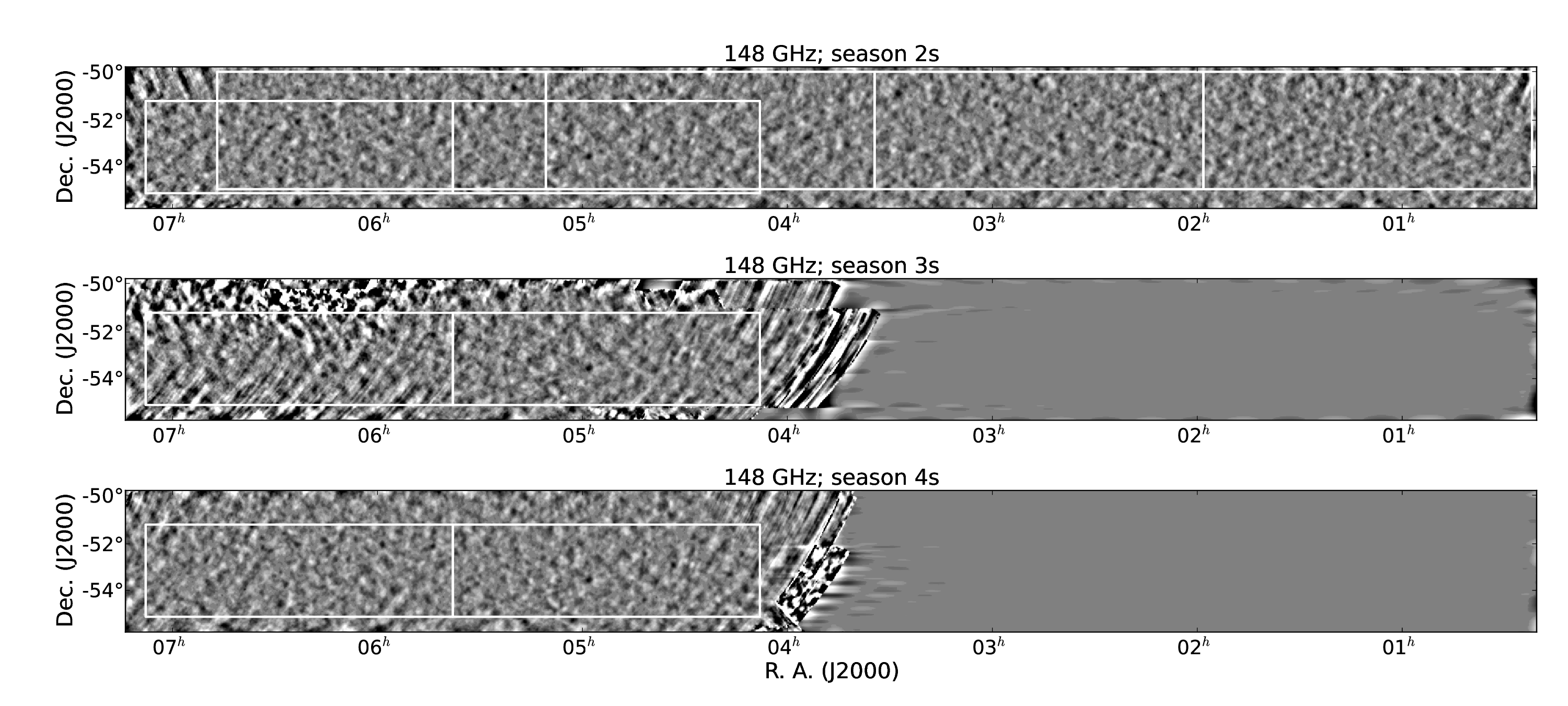}
\caption{Southern maps (\actS) made from 2008 (top panel) 2009 (middle panel) and 2010 (bottom panel) \arone\  observations  filtered to emphasize modes in the range $\ell=500-2500$. The four data splits were co-added to make this plot.  Also delineated are the patches used for computing power spectra.
 The smaller two patches common between the three maps are used to compute cross-season cross-power spectra. The four larger patches for season 2sf are 
 used to compute the full footprint 2008-only cross-power spectrum.    Areas of large noise or stripes are heavily down weighted in the analysis. The color scale is the same as Fig~\ref{fig:equMaps}. \label{fig:southMaps} }
\end{center}
\end{figure*}

\begin{figure}[htbp]
\begin{center}
\includegraphics[width=\columnwidth]{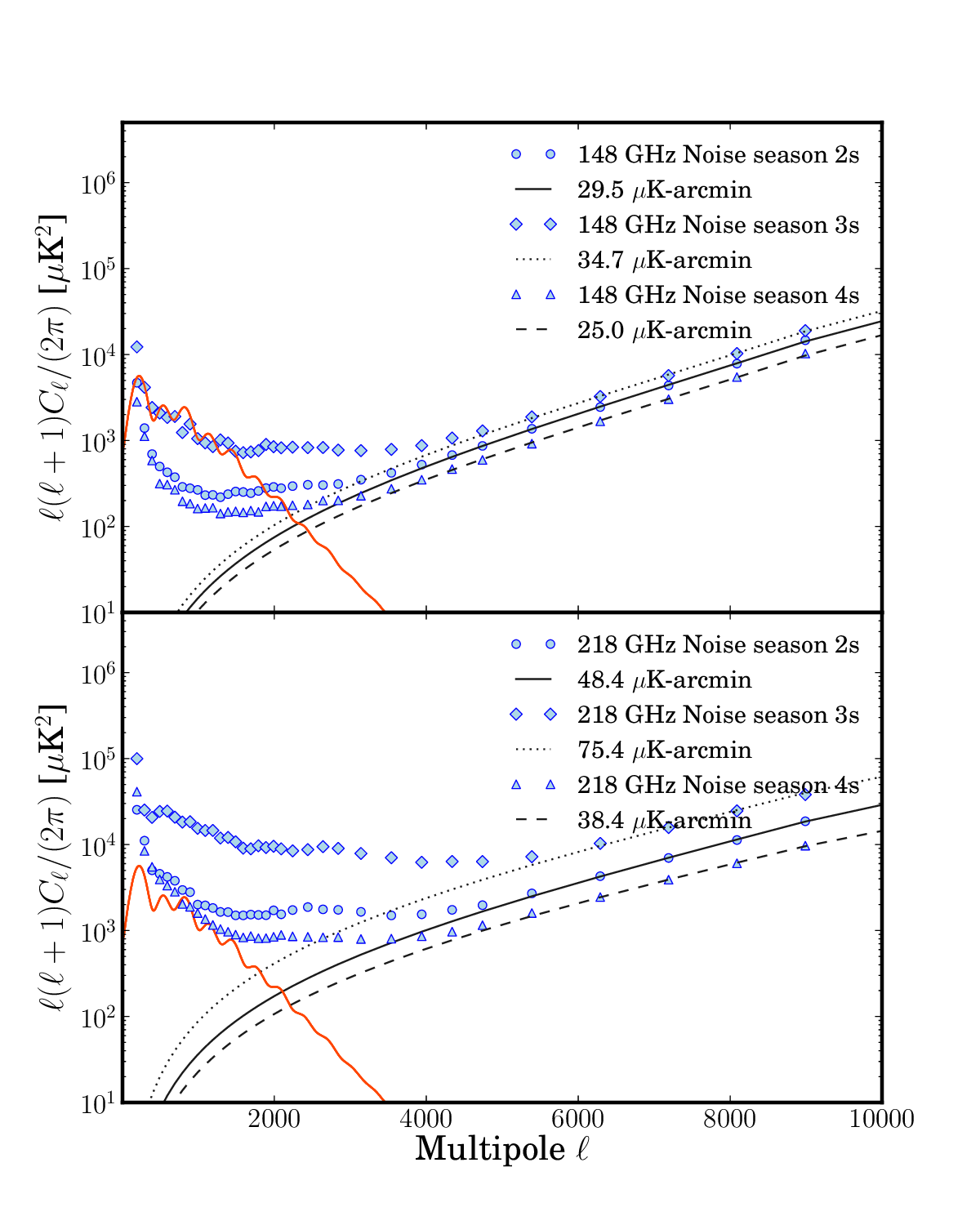}
\caption{Noise spectra for each seasons for the \actS\ maps  for \arone\ (upper panel) an \artwo\ (lower panel).   
The red solid line shows the CMB-only spectrum.
Season 3s is significantly  noisier than the other two seasons.  Note that the combination of seasons 3s and 4s is more sensitive
 than season 2s which was used in \citetalias{das/etal:2011} and \citet{dunkley/etal:2011}. \label{fig:southNoise}}
\end{center}
\end{figure}

The observations made on the southern sky across various seasons had  different footprints,  requiring a somewhat involved strategy for efficiently 
computing the power spectrum. Filtered versions of  various season maps are shown in Fig.~\ref{fig:southMaps}. The largest coverage was obtained
in the 2008 season (the same area on which \citetalias{das/etal:2011} was based). When computing the power spectrum within the  2008 data set, we used four large patches collectively covering 292 deg$^2$ (we refer to this full footprint as ``season 2 south full'' or season 2sf in short). 
  For computing 
the power spectrum within the  other two seasons, as well as to compute the cross-power spectra between any pair of the three seasons it was necessary to define another set of two patches (shown
by the smaller contiguous rectangles in Fig.~\ref{fig:southMaps}) that had a common footprint across the seasons.  As discussed in Section~\ref{sec:powspec}, care was taken not to double count information while combining the different spectra. The noise power spectra of each of the season maps 
are displayed in Fig.~\ref{fig:southNoise} against the CMB-only theory. The season 3s  map is mostly noise dominated  on all scales in either frequency -- we 
 keep this season in our analysis to tease out information  from cross-season spectra, but the season 3s-only spectrum is heavily downweighted in our 
 likelihood. 

\subsection{From Time-Ordered Data to Maps}

  \begin{figure}
\includegraphics[width=\columnwidth]{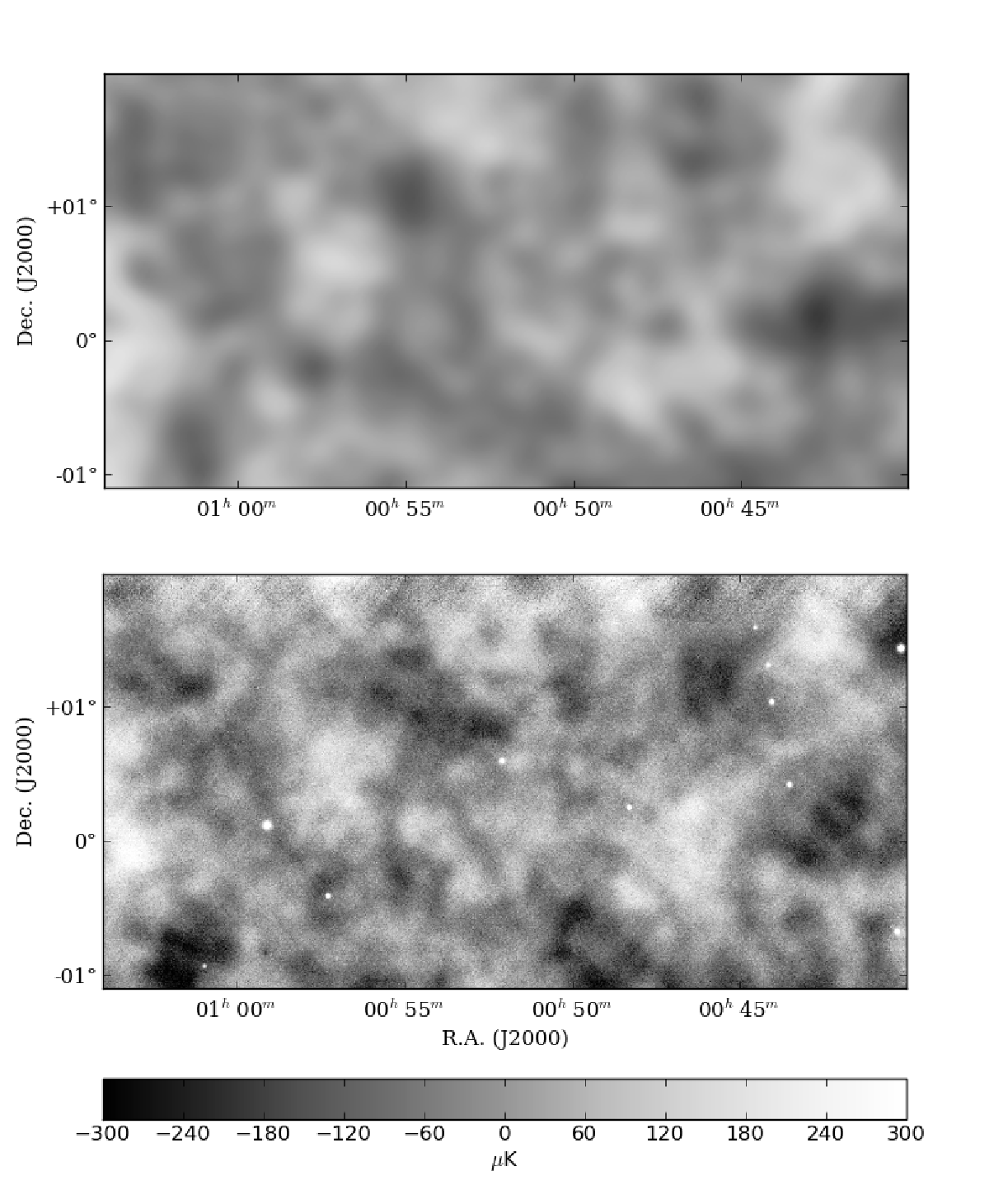}
\caption{Comparison of a sky patch from the WMAP 7-year 94 GHz map \citep{jarosik/etal:2011} (top) with the map of the  same region made from ACT \arone\ (bottom) observations  (co-added across seasons).  
All maps have been
high-pass filtered with a $\cos^2 \ell$-like filter that goes from 0 to 1 for $100 < \ell  <  300$. Agreement between the CMB features in the two maps is clear by eye. \label{fig:ACTvsWMAP}}
\end{figure}

  \begin{figure*}
\includegraphics[width=2\columnwidth]{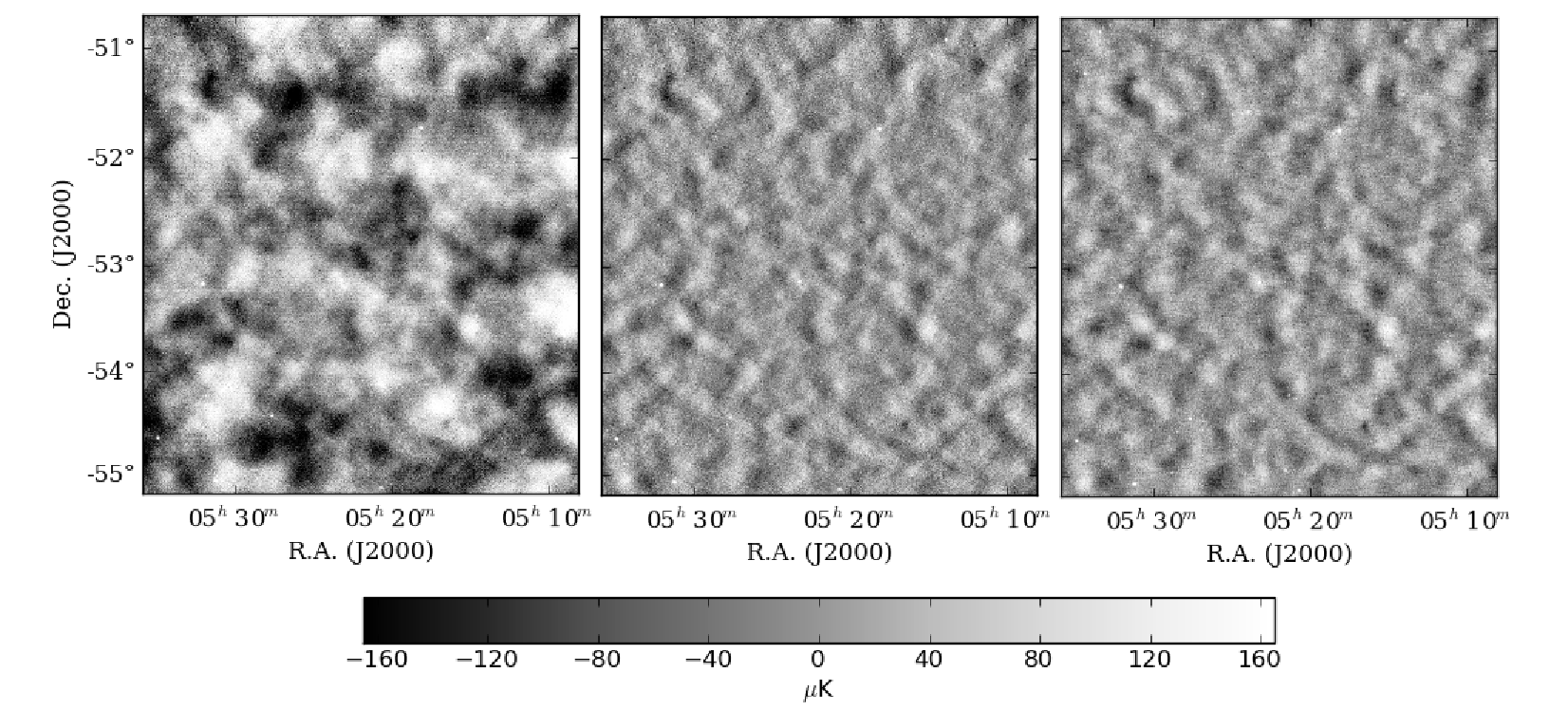}
\caption{Side by side comparison between the ACT map (co-added across seasons) and the SPT map for the same region of the sky. 
The left panel shows the ACT map
high-pass filtered with a $\cos^2 \ell$-like filter that goes from 0 to 1 for $100 < \ell  <  300$, and the center and right panels show the ACT and SPT maps respectively under the same high-pass filter used in the SPT data release \citep{schaffer/etal:2011}. Agreement between the CMB features in the two maps is clear by eye. It is noteworthy that the instrumentation, scan strategy, and analysis methods for these two experiments are completely different. \label{fig:ACTvsSPT}}
\end{figure*}

Details of the map-making procedure starting from time-ordered data  can be found in \citet{dunner/etal:prep}. 
The maps were produced using the algorithm described in \citet{dunner/etal:prep},
and  the 148 GHz 2008 data are identical to the maps presented
there.  We present a short summary of the mapping
procedure here.  Before mapping, we reject any detector timestreams
that have too many spikes (such as cosmic ray hits) identified in the
data.  The cuts	were calculated	with differing sensitivites in the
spike finder for different bands and seasons; the threshold for	each
band/season is set to correspond to roughly the	same fraction of data
rejected because of spikes.  The thresholds are 11/9/6 spikes per
10-15 minute timestream	for the	218 GHz	2008/2009/2010 data, and
9/11/11 for the	2008/2009/2010 148 GHz data.  We then interpolate
across gaps in the remaining detector timestreams and deconvolve the
effects of the detector time constants and a (known) filter from the
readout electronics.  Next we remove an offset from each detector and
a single slope common across the array.  We then estimate the noise as
described in \citet{dunner/etal:prep}, using a model that finds correlations
across the array, and measures the power spectra of those correlations
in frequency bins and the power spectra of the individual detectors
after the correlations are removed (the dominant correlated signal is
a common-mode atmosphere signal, but both higher order atmosphere
signals and electronic noise produce correlated noise across the
array).  We then use a preconditioned conjugate-gradient (PCG) iterative
algorithm to solve for the maximum-likelihood maps. At the same time we
solve for the values of the timestreams in regions where the data have
been cut out.  The cut samples are assumed to be decoupled from the sky;
 the solution is effectively using the noise model to interpolate across the gaps.
 We do this procedure twice; the second time we subtract off
the first solution from the timestreams to avoid any signal in the
data from biasing the noise estimation.  We find that the maps are
typically unbiased to better than one part in $10^{-3}$ and in all
cases the transfer function is much smaller than the beam error in the
$\ell$ ranges we use for science and calibration.  Simulations show
that the maps typically	converge within	a few hundred PCG iterations.
\subsection{Beam Transfer Functions}

The beams are estimated independently for each array and season
\citep[in preparation]{hasselfield/etal:prep} from observations of Saturn following a
procedure similar to the one described in \cite{hincks/etal:2010}.
Radial beam profiles from the planet maps are transformed to Fourier
space by fitting a set of basis functions whose analytic transform is
known.  The fit yields the beam transform as well as a covariance
matrix following a procedure similar to that discussed in \citetalias{das/etal:2011}. 
 The transform is subsequently corrected for the mapper
transfer function, the solid angle of Saturn, and the difference in
Saturn's spectrum compared to the CMB blackbody spectrum.  Because any
location in the ACT CMB maps contains data from many different nights,
the effective beam in the maps is broadened relative to the
planet-based beam due to pointing variation from night to night.  This
pointing variation ($\simeq 6\arcsec$)  is modelled as having a two-dimensional Gaussian distribution,
and the standard deviation is measured by comparing the shape of the beam obtained from  stacked
radio sources to the planet-based beam transform.  The error in the
beam due to the pointing correction is included in the final beam
covariance matrix.  The covariant error in the beam is obtained after fixing the normalization 
of the beams at $\ell = 700~ (1500)$ for the 148 GHz (218 GHz) array.  The calibration error 
is thus separated from the covariant error due to beam shape uncertainty, which is 0 by construction at $\ell = 700~ (1500)$.

\subsection{Comparison with WMAP and  South Pole Telescope Maps}
Fig.~\ref{fig:ACTvsWMAP} shows a comparison between  the WMAP 7-year 94 GHz map \citep{jarosik/etal:2011}, and the \arone\ 
and \artwo\ ACT maps on the same region of the sky.  On the one hand, it exemplifies how our map-making pipeline faithfully 
reproduces all the large-scale CMB features seen  in the WMAP map, and on the other hand it portrays the significantly
higher resolution afforded by ACT over WMAP.  
This figure is a visual representation of the fact that the transfer function in our maps are unity down to small angular scales ($\ell \simeq 300$) 
or large spatial scales ($0\fdeg6$) -- this proves highly beneficial for calibrating our maps against the WMAP maps, as discussed later. 

We also compare ACT maps with publicly released maps from the South Pole Telescope \citep{schaffer/etal:2011} in the region
of overlap in the southern strip.   Fig.~\ref{fig:ACTvsSPT} shows a side by side comparison between the ACT map co-added across 
all seasons and the SPT map on the same region of the sky. In the middle panel, the ACT map has been filtered with the  same filter 
used for the SPT map. The similarity between the two maps is clear by eye, and speaks to the high quality of both measurements. 
A comparison of this figure with Fig.~13 of \cite{dunner/etal:prep}  (which used only 2008 observations) shows the improvement in the
  noise properties  of the ACT map from the combination of multi-season observations. \par

\begin{figure}
\centering
\includegraphics[width=\columnwidth]{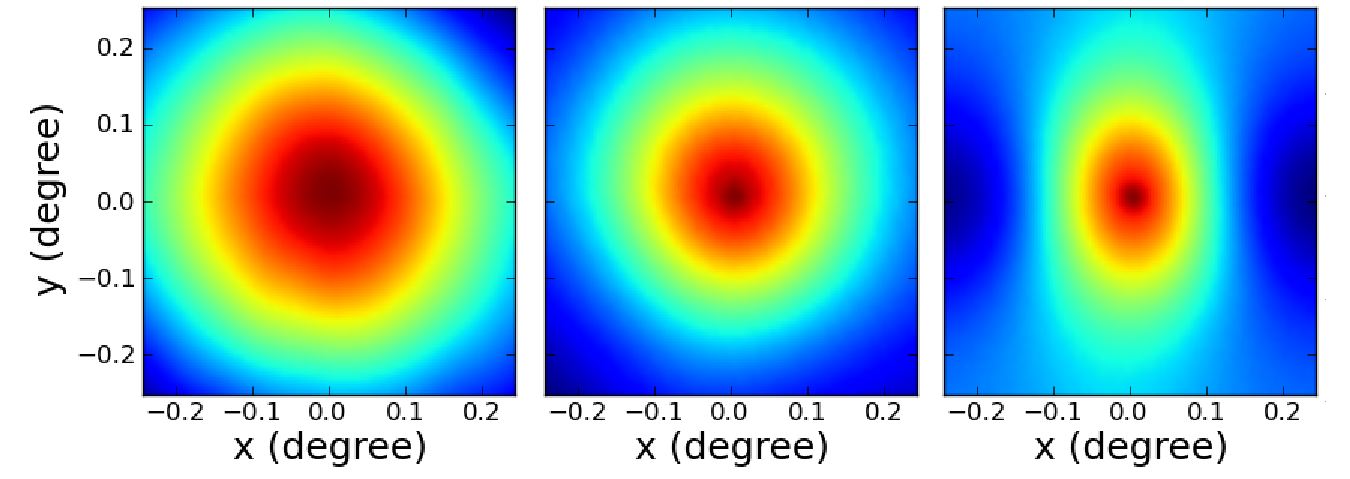}
\includegraphics[width=\columnwidth]{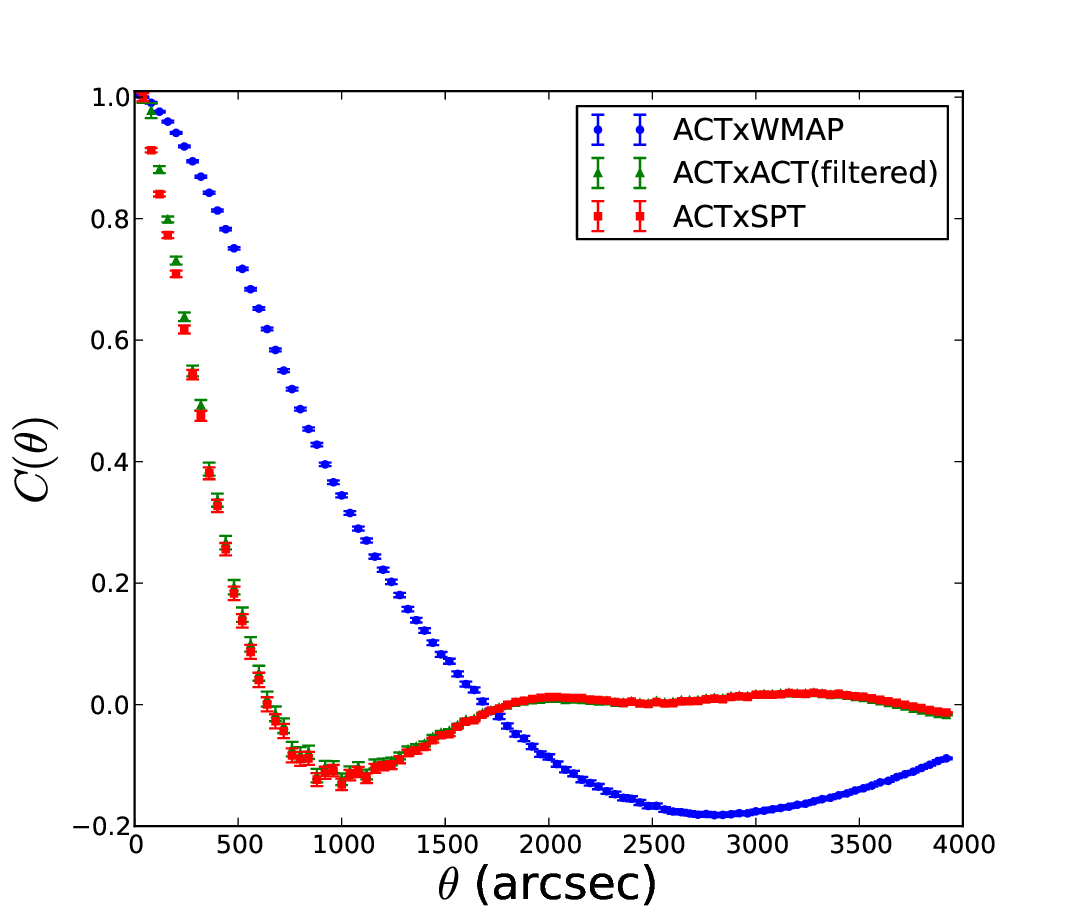}
\caption{{\emph{Top:}} Two-dimensional cross-correlation function in real space  between WMAP and ACT \arone\ map (left panel), 
between ACT \arone\ and ACT \artwo\ maps (middle panel),  and between the  ACT \arone\  and the SPT 150 GHz map from \citet{schaffer/etal:2011}. 
Note that the ACT-SPT cross-correlation is anisotropic as the SPT map is  filtered in the  $\ell_x$ direction to suppress modes below $\ell_x\lesssim 1200$. {\emph{Bottom:}} One-dimensional binned version of  real space correlation functions.  The blue dots represent the 1D ACT \arone\ $\times$ WMAP correlation, while the green triangles represent  the 1D auto-correlation of ACT \arone\  maps after the SPT filter has been applied to them.  Finally, 
the red squares denote the 1D cross-correlation of the ACT \arone\ map with the SPT map. \label{fig:ACTxSPT}}
\end{figure}

  Using the techniques described in \citet{hajian/etal:2011}, we also studied  the two-dimensional real-space correlation function  (cf.~equation 3 of \citealt{hajian/etal:2011})  between the ACT and the SPT maps. Both the 2D cross correlation and
   the binned 1D  version are shown  in Fig.~\ref{fig:ACTxSPT}. 
The cross-correlation is anisotropic as the SPT map is filtered in the  $\ell_x$ direction to suppress modes below $\ell_x\lesssim 1200$, but 
the overall agreement between the two experiments is excellent.

\section{Calibration}
\label{sec:calib}

The final map calibration is performed in two stages: first, the 148 GHz map from the 
lowest-noise season is calibrated against the WMAP sky map, and then the 148 GHz
maps from the remaining seasons and 218 GHz maps from all seasons are calibrated
against the WMAP-calibrated 148 GHz map. 

\subsection{WMAP Calibration}
ACT map-making  and observing strategies result in maps with unbiased large-angle modes that can be compared to WMAP maps of the same region. The maps are cross-linked, i.e. every point in the survey has been observed during both its rising and setting. The cross-linked data are fed to  a  map-making pipeline described in  \citet{dunner/etal:prep} that allows the reconstruction of all modes in the map  without biasing the large-angle modes. The transfer function of the maps is unity  to better than 1\% at angular scales corresponding to $\ell > 300$  \citep{dunner/etal:prep}.  We calibrate the 148 GHz ACT maps directly to WMAP 7-year 94 GHz maps \citep{jarosik/etal:2011} of the identical regions using the cross-correlation method described in \citet{hajian/etal:2011}. By matching the ACT-WMAP cross-spectrum to the ACT power spectrum and the WMAP 7-year power spectrum \citep{larson/etal:2011} in the range $300 < \ell < 1100$,  we calibrate the  148 GHz ACT spectrum to 2\% fractional temperature anisotropy uncertainty \citep{hajian/etal:2011}.  Calibration to WMAP is done on the deepest seasons for both \actE\ and \actS\ strips, which correspond to season 4e (2010 observing season) and season 2sf (2008 observing season) data respectively. These calibrated maps are used as references to calibrate other seasons as described below. 
\subsection{Relative Season Calibration} 
Once the deepest \arone\  season has been calibrated with respect to WMAP, we cross-correlate that map with a
 \arone\ map from another season,  take the ratio of the cross-season cross-power spectrum to the in-season 
 cross-power spectrum, and fit for a calibration factor. For example, on the equator, the \arone\  season 4e  map is calibrated 
 with respect to WMAP. We then compute the ratio $\cl^{4e \times 4e}/\cl^{3e \times 4e}$ to estimate the calibration for the season 3e \arone\ map. This  internal method lets us use a much wider range of angular scales (we use an $\ell$ range of width 2000 starting at $\ell=500$  for \arone\, and  $\ell=1000$ for \artwo)  than possible with  WMAP.  
 Using this method,   we achieve the following relative calibration uncertainties (expressed as $\sigma_{X-Y}$   for season X calibrated 
 against season $Y$) : $\sigma_{3e-4e} \simeq 0.7\%$, $\sigma_{3s-2s} \simeq 3\%$, and $\sigma_{4s-2s} = 3\%$ for \arone, and 
  $\sigma_{3e-4e} \simeq 2\%$, $\sigma_{3s-2s} \simeq 9\%$, and $\sigma_{4s-2s} = 4\%$ for \artwo\ maps. Note that for the  \actS\ 
  season 3s maps the calibration uncertainties are higher, as is expected from the fact that this season is largely noise dominated 
  on most  scales  (see  Fig.~\ref{fig:southNoise}). The internal spectrum from this season gets highly downweighted in our likelihood.  
 To tie together the \arone\ and \artwo\ internal calibrations,  we  finally calibrate the best season \artwo\ map with respect to its \arone\ counterpart, 
 achieving $\sigma_{2s ( 218) - 2s (148)} \simeq 1.3\% $,  and $\sigma_{4e ( 218) - 4e (148)} \simeq 1.7\% $.  This gives the overall 
 calibration of the reference \artwo\ map as  $ 2.4\% $ for \actS\ and $2.6\%$ for \actE.

\section{Temperature Power Spectrum Analysis}
\label{sec:powspec}
The power spectrum analysis methods used here are essentially the same as in \citetalias{das/etal:2011}, the major 
modifications being due to the smaller extent of the \actE\ maps in the declination direction compared to \actS, and the multi-season nature of the spectra. 
\subsection{Preprocessing of maps}
We follow \citetalias{das/etal:2011} and apply a high-pass filter to the maps to suppress the large angular scale modes ($\ell<500$)
 which are not as well constrained as others, and can bleed power into the smaller-scale modes. Next, we prewhiten the maps  through real space operations 
 as in  \citetalias{das/etal:2011}. 

\subsection{Data Window}
Each patch is then multiplied with a data window before the power spectrum is computed.   The window is a product of three components:
a point source mask, an apodization  window, and the \nobs\ map giving  the number of observations in each pixel.   
The point source mask is further described in Section~\ref{ssec:PS}. 
To simplify the application, we create a single \nobs\ map per patch by adding the individual \nobs\ maps from 
all the splits involved (four splits for the single frequency spectrum, and the 8 splits for the cross-frequency or cross-season spectrum), and apply this 
as a weight function.  This essentially downweights the poorly observed regions of the patch.  An additional step is applied to the 
\actE\ patches. Since the \actE\ strip is only 3 degrees wide, the absolute Fourier space resolution in the declination direction is  $\Delta \ell_y = 120$. 
This leads to instability in the mode-mode coupling calculations due to poor sampling of power in the Fourier space. To remedy this, we extend the patches in the declination direction by adding a 0.7-degree-wide-strip of zero valued pixels on either side, such that the final declination width of the zero-padded patch is 4.4 degrees.   To minimize ringing from the edges of the patch, we also
apply an apodization window which is generated by taking a top-hat function that is unity in the center and zero over 10 pixels at the edges of the original patch, 
and convolving it with a Gaussian of full width at half maximum  of  2\farcm 5 for \actS\ and 14\farcm 0 for \actE . Another addition to the pipeline for the \actE\ patches is the application of a Galactic dust mask  (see Section~\ref{ssec:GD}). Monte Carlo simulations  demonstrate  that we retrieve an unbiased
estimate of the spectrum with these additions to our well-tested pipeline.  \par
%\tbd{this prompts larger bins than D11}
\subsection{Binning of the power spectrum}
\label{ssec:binning}
It is important to note that the narrow inherent width of the \actE\ strip, as well as the smaller dimensions of the mulitseason \actS\ 
patches  prompted us to adopt wider bins for the bandpowers than were used in \citetalias{das/etal:2011}. Most notably, over the acoustic peaks ($\ell<2000$) the bins used have a width $\Delta\ell=100$  as opposed to $\Delta\ell=50$ of  \citetalias{das/etal:2011}.  This choice is
motivated by the fact that with the finer binning, adjacent bins remain significantly correlated for the \actE\ spectrum. In addition, evaluating the covariance
using   full  end-to-end  Monte Carlo simulations is  prohibitively expensive given the iterative nature of our map-making process. Conversely,  more 
tractable  approximations might not be good enough to provide the  precision deserved by the high quality of the data.  With the larger bins we have verified
that the bin-to-bin correlations never exceed  10\% and are much smaller than 10\%  for most bin pairs, allowing us to treat the bandpowers  as statistically independent (cf. Section~\ref{ssec:altBins}) . 

\begin{figure*}[htbp]
\begin{center}
\includegraphics[scale=0.34]{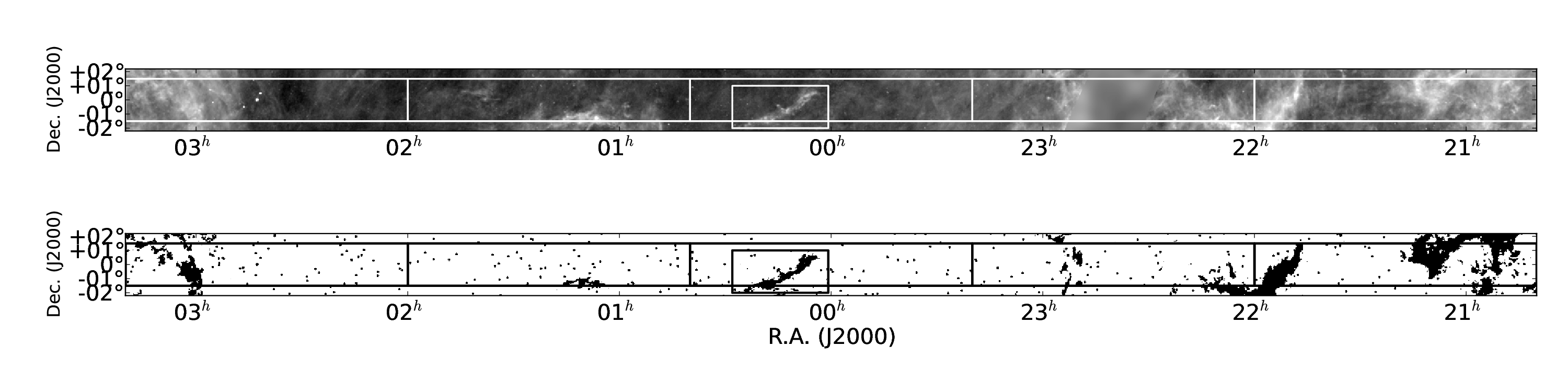}
\caption{\emph{Top:} The IRIS  100 \micron\ map  on the \actE\ strip (arbitrary color scale). \emph{Bottom:} The
equatorial dust mask  based on the IRIS map flux cut as described in the text. The small box near right ascension of \hmin{00}{14} shows the ``seagull''-like 
structure that is additionally masked out even though some pixels fall below the flux cut. \label{fig:equDust}}
\end{center}
\end{figure*}

\subsection{Cross-Season Cross-Spectrum Estimation}
To obtain unbiased estimates of the cross spectrum we  follow the same steps as enumerated in Section 3.6 of \citetalias{das/etal:2011}, 
which involve deconvolving  a mode-coupling matrix that accounts for the effects of beam profile, prewhitening, filtering, pixelization, and windowing.
For same-season cross spectra, the combinatorics are exactly the same as in \citetalias{das/etal:2011}: we compute 6 cross spectra per patch for the single-frequency spectrum (the 4 ``splits'' in time giving rise to the 6 cross spectra), and 12 for the cross-frequency spectrum (avoiding crossing the same splits that contain data from the same nights).  For cross-season 
spectra we combine all 16 cross-season cross-spectra, as each  split from one season has independent noise from the splits in the other season. For each frequency and season pair,   the cross-spectra from the patches are combined with inverse variance weighting. This results in  a  set of three bandpowers 
$\{C_b^{3e\times 3e}, C_b^{3e\times 4e}, C_b^{4e\times 4e}\}$ for \actE\, and  a set of  six  bandpowers  $\{C_b^{2s\times 2s}, C_b^{2s\times 3s}, C_b^{2s\times 4s}, C_b^{3s\times 3s}, C_b^{3s\times 4s},C_b^{4s\times 4s}\}$ for \actS, for each of the two same-frequency pairs $\ara\times \ara$ and $\arb \times \arb$. For the cross-frequency spectra $\ara\times\arb$,  where  $C_b^{3e (\ara) \times 4e(\arb)} $ is distinct from    $C_b^{4e (\ara) \times 3e(\arb)} $, we get  a set of four bandpowers for \actE\ and nine bandpowers  for \actS. These add up to a total of 10  cross-power spectra for \actE\ and 21 cross-power spectra for \actS\ that enter the likelihood separately with their individual 
bandpower covariance matrices. Additionally, there are six cross-power spectra coming from the full-footprint 2008 \actS\ map (2sf), which is added, with proper attention to the overlap between the s2f and s2 patches. 

\subsection{Bandpower Covariance}
For each cross-power spectrum above, we evaluate a bandpower covariance matrix:
\ba
\nonumber \Theta^{(\alpha A\times \beta B);( \gamma C\times \tau D)}_{b b} &=& \left \langle\left(\hat C_b^{\alpha A\times \beta B }-\ave{\hat C_b^{\alpha A\times \beta B}}\right) \right.\\ 
&& \times \left. \left( \hat  C_b^{\gamma C\times \tau D} - \ave{\hat C_b^{\gamma C\times \tau D}} \right)\right\rangle
\ea
where we used Greek indices $\alpha$, $\beta$, etc. to denote the seasons  3e, 2s, etc. and the uppercase Roman numerals $A$, $B$, etc. to denote frequencies.  The analytic expression of the general term of this covariance matrix $\Theta^{(\alpha A\times \beta B);( \gamma C\times \tau D)}_{b b} $  is discussed in the Appendix.  The total covariance matrix is a sum of two  terms:  a  sample covariance  part accounting for the fact that different seasons of observation at different frequencies are observing the same CMB modes and that some of our cross spectra have common noise,  and another part coming from the covariance of uncertainties in the  determination of the beam profile.  The covariance matrix is computed analytically and checked against Monte Carlo simulations described in Section~\ref{sec:sims}. This total covariance matrix is used in defining the likelihood function \citep{dunkley/etal:prep, sievers/etal:prep} when determining cosmological parameters. \par 
Along with the covariance matrices, we also generate bandpower window functions which convert a theoretical power spectrum into a band power: $C_b^{\rm{th}} = \sum_\ell~ B_{b\ell} C_\ell^{th}$. Due to different 
geometry and noise properties of the \actE\ and \actS\ patches, two separate window functions $B_{b\ell; ACT-s}$ and $B_{b\ell; ACT-e} $ 
are evaluated for the south and the equator.

\subsection{Combining Multi-Season Spectra}

For the purpose of parameter estimation, we keep the \actE\ and \actS\ spectra and covariances for each season separate in the likelihood \citep{dunkley/etal:prep}. For display purposes and for visual comparison with other datasets we  combine the spectra from different seasons (separately 
for equator and south) using
inverse variance weighting:
\ba
\tilde{C}^{A\times B}_{b}= \frac{\sum_{\alpha,\beta}  (\Theta^{-1})^{Ê(\alpha A\times \beta B);Ê(\alpha A\times \beta B)}_{bb} C_{b} ^{(\alpha A \times \beta B)}}{\sum_{\alpha,\beta}  (\Theta^{-1})^{Ê(\alpha A\times \beta B);Ê(\alpha A\times \beta B)}_{bb} } .
\ea
As discussed in \citet{dunkley/etal:prep} the amplitudes of the Galactic cirrus contributions to the \actE\ and \actS\ maps are different. Therefore, before 
combining the \actE\ and \actS\ spectra  obtained above, we subtract the best-fit cirrus component (see \ref{ssec:GD} for more details) from the \actE\ and \actS\ spectra, and then combine them using inverse variance weighting. The multiple levels of cross-correlation used in computing the power spectrum help ensure that potential peculiarities in the observation that are located in time or space do not propagate to the final power spectrum.

\section{Foregrounds}
In the \arone\ and \artwo\ bands the main foregrounds are emission from point sources and diffuse Galactic dust, which we treat with the application 
of masks as described below. In the companion papers \citet{dunkley/etal:prep} and \citet{sievers/etal:prep} we also consider and constrain contributions
from thermal and kinetic Sunyaev Zaldovich effects, clustered  and Poisson-like infrared point sources, radio sources,  and a residual Galactic cirrus component. 

\label{sec:foreg}
\subsection{Galactic Dust \label{ssec:GD}}
We detect  a significant contribution from Galactic cirrus in our \actE\ maps, especially at \artwo. We employ a two-step approach 
for dealing with Galactic cirrus in the \actE\ maps using 100~$\micro$m dust maps from IRIS \citep{miville-deschenes/lagache:2005} 
as the reference.  The first step is motived by the observation that most of the dust contamination in the equatorial 
power spectrum comes from the regions corresponding to bright clustered structures in the dust map. Therefore, we generate 
a dust mask by identifying and setting to zero all pixels above a flux density of 5.44  MJy/sr as well as pixels that fall inside significantly clustered structures 
such as the ``seagull''-like structure near right ascension of \hmin{00}{14} shown inside the box in Fig.~\ref{fig:equDust}.  This dust mask is multiplied by the point source 
mask described below to generate the final mask that is applied to the data. 

The second step of the dust treatment is generating a model of the 
residual dust contamination after the application of this mask, and then to inform the parameter estimation pipeline with reasonable priors on this model (the residual model amplitude is fitted and marginalized over when constraining cosmological parameters, as described in \citealt{dunkley/etal:prep}).  The model is constructed as follows. Following \citet{hajian/etal:2012} we perform a multicomponent fit to the auto-power spectrum of the IRIS map after application of the dust mask described above. The components include  a power law term for the residual Galactic cirrus, a Poisson shot noise term,   a term representing the clustered component of  infrared emission, and a white noise term to describe the instrument. The Galactic cirrus component is  modeled as $\cl^{\rm{cirrus}} = A_{\rm{cirrus}} ~ \ell^{-2.7}$, where the value of the power law index appears to be a good fit to Herschel  observations of cirrus \citep{miville-deschenes/etal:2010, bracco/etal:2011}, as well as  the cross correlation between ACT and BLAST \citep{hajian/etal:2012},  and  that between ACT  and IRIS maps.  This fitting procedure provides us with an estimate of the amplitude $A_{\rm{cirrus}}$ separately for the \actE\ and the \actS\ map 
%(both the full 2008 season or 2sf, and the multi-season overlap region or s2s) 
footprints. Next,  we cross-correlate the ACT maps with the IRIS template to evaluate the dust coefficient 
$A_d = \cl^{\rm{ACT}\times\rm{IRIS}}/\cl^{\rm{IRIS}\times\rm{IRIS}}$ for each frequency and each sky region.  
 Finally, the cirrus contribution to the ACT power spectrum can be expressed as $\cl^{\rm{gal}}  = A_{\rm{cirrus}} A_d^2 ~  \ell^{-2.7}$ or  expressed in terms of a rescaled amplitude  at $\ell_0 = 3000$:  ${\cal B}_\ell \equiv \ell^2 C_\ell /(2\pi) = a_{\rm{g}}  \left(\ell/{\ell_0}\right)^{-0.7}$, 
 where we have defined $a_{\rm{g}} \equiv  A_{\rm{cirrus}}~ A_d^2 ~\ell_0^{-0.7}/(2\pi) $. 

 \begin{deluxetable}{ccccc}
 \tablewidth{0pc}
 \tablecolumns{5}
\tablecaption{ Coefficients for  Galactic cirrus model\label{table:dustModel}}
\tablehead{
\colhead{Region} & \colhead{$A_{\rm{cirrus}}$} & \colhead{Frequency}  & \colhead{$A_{d}$ \tablenotemark{a} }& \colhead{$a_g $}\\
\colhead{} & \colhead{MJy$^2$} & \colhead{GHz}    &  \colhead{\micro K/MJy} & \colhead{}}
\startdata
\hline
&&&& \\
 \actS & 9.95  & 148 & 8.65  & 0.4  \\
 &   & 218  & 30.0  & 5.2 \\
\hline
&&&& \\
\actE & 17.9  & 148 & 8.65  & 0.8 \\
 &  & 218  & 30.0  & 9.4 
\enddata 
\tablenotetext{a}{we use a common dust coefficient for equator and south}
\end{deluxetable}
 
\begin{figure}[htbp]
\begin{center}
\includegraphics[width=\columnwidth]{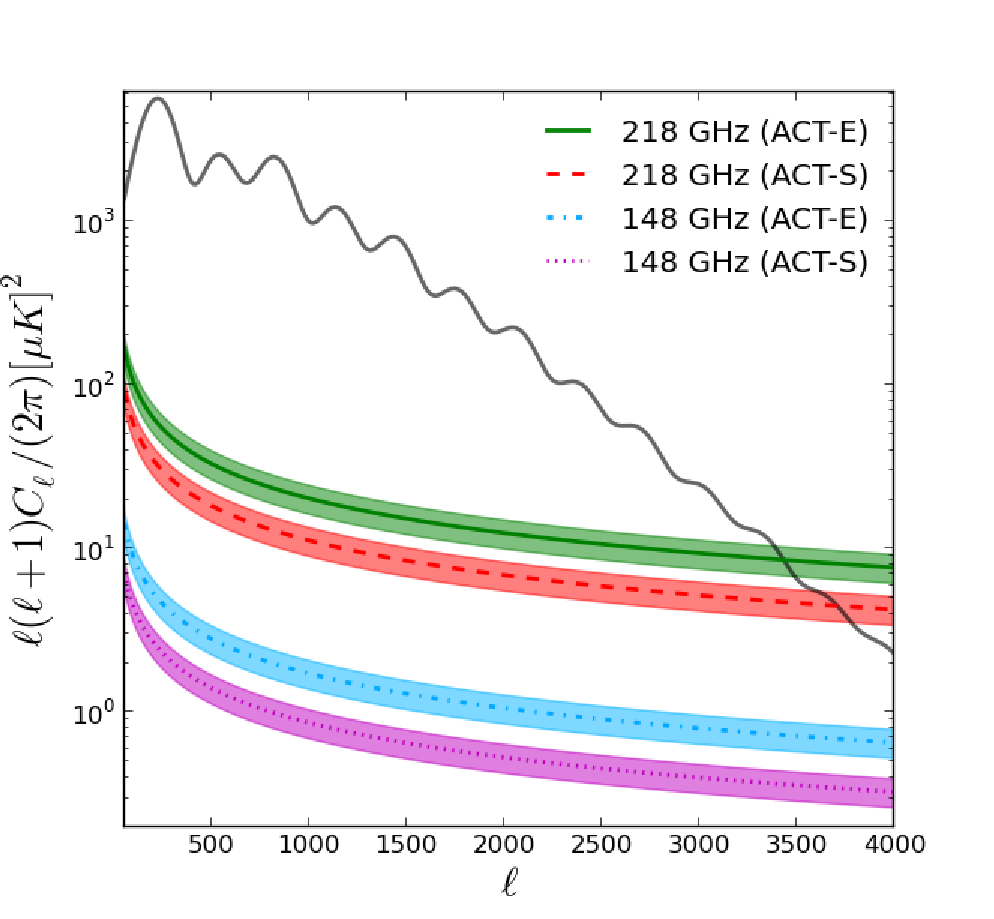}
\caption{Galactic cirrus contributions to the power spectrum modeled as a power law as  described in the text, for each frequency and region of the sky.  
The curves correspond to  the best fit amplitudes obtained by cross-correlating ACT maps with cirrus component of the IRIS maps, and the band around them represent the uncertainty adopted as a prior in the  ACT likelihood 
 as  discussed in \citet{dunkley/etal:prep}.  \label{fig:dustFig}}
\end{center}
\end{figure}

 The various model parameters obtained from the fitting method above are displayed in Table~\ref{table:dustModel}.  There
 is roughly twice as much dust in the \actE\ region as in \actS, but at $\ell=3000$ and for \arone\ it is less than 3\% of the CMB signal. These values 
represent a  frequency scaling  consistent with the early release results from the  Planck satellite \citep{planck/etal:2011}, and can be compactly 
written in flux units as ${\cal B}_\ell^{ij} = a_g (\ell/\ell_0)^{-0.7} (\nu_i \nu_j /\nu_0^2)^\beta$ $\micro$K$^2$ with $\beta = 1.8$,   $\nu^i$ and $\nu^j$ 
the two frequencies being  crossed, and $\nu_0 = $\arone.  Based on the scatter observed in these central values as well their variation depending on 
whether the clustering term is included in the fit, we adopt priors of $a_{gs} = 0.4 \pm 0.2$ and $a_{ge}= 0.8\pm 0.2$  for \actS\ 
and \actE\ respectively. The fitted models and priors are illustrated in Fig.~\ref{fig:dustFig}.  

\subsection{Point Sources \label{ssec:PS}}

Point sources are the main astrophysical foreground for the 148 and 218 GHz  bands.  At these frequencies, point sources are typically either radio-loud AGN or dusty star-forming galaxies.   Most of the bright sources are AGN, while most of the dusty star-forming galaxies lie below the detection threshold of our survey. Point sources must be identified and masked before the power spectrum is computed so as not to add power to the cosmological signal.  We have identified sources using a matched filter algorithm \citep[e.g.,][]{tegmark/de:1998}.  We mask data within a 5\arcmin\ radius around all sources detected down to 15 mJy in either band.  The residual power contributed to the power spectrum from unmasked sources below our detection threshold is expected to be $2.9\pm0.4 ~\micro$K$^2$ at $l = 3000$ (Gralla et al., in preparation).
 For details about the point source detection algorithm we used, and catalogs for the south 2008 \arone\ data, see \citet{Marriage/etal:2011}.  Catalogs for the remaining data set will be published in subsequent papers.

%\subsection{SZ and correlated IR background}

\section{Simulations}
\label{sec:sims}
We ran a set of Monte Carlo (MC) simulations in order to validate the analytic prescription for the uncertainties on the  cross-season cross-frequency  power spectra, and to investigate bandpower covariance and possible biases in the pipeline. 
As our map-making procedure is iterative, it is prohibitively expensive to run a large set of end-to-end simulations that would capture all aspects of the map-making pipeline, and  the noise characteristics and correlations in the actual data set. 
Instead, following \citetalias{das/etal:2011},  we  generate signal maps as Gaussian random realizations from a power spectrum,  and add to each of them a  realization of a Poisson point source population, and a  simulated noise map generated from  the observed noise-per-pixel in the data maps. The details of the implementation are essentially the same as  in Section 4 of \citetalias{das/etal:2011} with special care  taken so that signal  realizations are properly correlated across different seasons and  footprints.  For each season and each frequency, we generate 960 signal+noise maps (four splits for each frequency), and for each realization we compute the power spectra in exactly the same way as we do for the data maps. From the large set of cross-power spectra obtained in this way we estimate the season-season 
covariance as well as the correlation between band powers. We find that in all cases, correlation between adjacent bins are insignificant at the 10\% level. \par
We evaluate the uncertainties in the band powers using an analytic prescription described in Appendix \ref{apdx:errorbars}. We verify the accuracy of 
these expressions by comparing  the predicted error bars with the scatter of  MC realizations. For isotropic white noise realizations with uniform weights, our expressions 
for multi-season multi-frequency error bars are good to better than a percent.  

%With the more realistic anisotropic noise seeded by the data noise  as described above, there are small corrections  to the analytic covariance, which we evaluate against the MC simulations. 

\section{Temperature Power Spectrum Results}
\label{sec:results}
\begin{deluxetable*}{cc|cc|cc|cc}
\tablewidth{0pt} 
%\tablecolumns{8} 
\tablecaption{Single frequency combined bandpowers provided for plotting purposes only.\\ ${\cal B}_b = \ell_b (\ell_b+1) C_b/2\pi $   (\micro\kelvin$^2$) \label{tab:spec_table} }
\tablehead{\multicolumn{2}{c}{} & 
\multicolumn{2}{c}{\arone} &
\multicolumn{2}{c}{\arone{} $\times$ \artwo} & \multicolumn{2}{c}{\artwo}\\
\colhead{$\ell$ range} & 
\colhead{central $\ell_b$} &
\colhead{${\cal B}_b$ }& 
\colhead{$\sigma({\cal B}_b)$ }& 
\colhead{${\cal B}_b$}& 
\colhead{$\sigma({\cal B}_b)$  }& 
\colhead{${\cal B}_b$ } &
\colhead{$\sigma({\cal B}_b)$ }
}
\startdata      
540 - 640 & 590 & 2267.4 & 114.3 & - & - & - & - \\ 
 640 - 740 & 690 & 1760.2 & 79.4 & - & - & - & - \\ 
 740 - 840 & 790 & 2411.2 & 97.0 & - & - & - & - \\ 
 840 - 940 & 890 & 1962.4 & 75.5 & - & - & - & - \\ 
 940 - 1040 & 990 & 1152.2 & 42.8 & - & - & - & - \\ 
 1040 - 1140 & 1090 & 1208.7 & 43.2 & - & - & - & - \\ 
 1140 - 1240 & 1190 & 1057.7 & 36.0 & - & - & - & - \\ 
 1240 - 1340 & 1290 & 743.1 & 25.7 & - & - & - & - \\ 
 1340 - 1440 & 1390 & 833.3 & 27.3 & - & - & - & - \\ 
 1440 - 1540 & 1490 & 683.0 & 22.0 & - & - & - & - \\ 
 1540 - 1640 & 1590 & 484.7 & 16.4 & 494.0 & 15.7 & 551.2 & 30.0 \\ 
 1640 - 1740 & 1690 & 403.1 & 13.3 & 400.6 & 12.6 & 458.5 & 26.9 \\ 
 1740 - 1840 & 1790 & 377.7 & 12.3 & 369.7 & 11.5 & 408.3 & 23.9 \\ 
 1840 - 1940 & 1890 & 266.8 & 9.3 & 272.7 & 9.3 & 327.0 & 20.3 \\ 
 1940 - 2040 & 1990 & 236.5 & 8.7 & 261.3 & 8.8 & 320.2 & 20.5 \\ 
 2040 - 2140 & 2090 & 229.2 & 8.1 & 226.6 & 7.7 & 274.7 & 17.9 \\ 
 2140 - 2340 & 2240 & 150.2 & 4.3 & 168.6 & 4.6 & 238.7 & 12.3 \\ 
 2340 - 2540 & 2440 & 109.2 & 3.5 & 128.1 & 3.8 & 199.8 & 10.7 \\ 
 2540 - 2740 & 2640 & 75.0 & 3.0 & 97.5 & 3.3 & 181.6 & 9.8 \\ 
 2740 - 2940 & 2840 & 63.7 & 2.9 & 86.0 & 3.2 & 181.9 & 9.6 \\ 
 2940 - 3340 & 3140 & 43.9 & 1.9 & 69.4 & 2.1 & 167.3 & 6.4 \\ 
 3340 - 3740 & 3540 & 35.7 & 2.0 & 65.2 & 2.1 & 183.6 & 6.5 \\ 
 3740 - 4140 & 3940 & 36.1 & 2.4 & 68.3 & 2.3 & 211.4 & 7.3 \\ 
 4140 - 4540 & 4340 & 34.4 & 2.8 & 75.5 & 2.7 & 245.7 & 8.3 \\ 
 4540 - 4940 & 4740 & 37.3 & 3.4 & 93.8 & 3.3 & 286.8 & 9.8 \\ 
 4940 - 5840 & 5390 & 50.5 & 3.2 & 109.8 & 3.0 & 355.8 & 9.5 \\ 
 5840 - 6740 & 6290 & 59.5 & 5.2 & 149.6 & 4.6 & 478.0 & 14.6 \\ 
 6740 - 7640 & 7190 & 81.6 & 8.7 & 177.9 & 6.6 & 564.4 & 19.9 \\ 
 7640 - 8540 & 8090 & 131.4 & 14.6 & 240.1 & 10.7 & 753.1 & 29.6 \\ 
 8540 - 9440 & 8990 & 133.0 & 25.5 & 265.3 & 16.6 & 878.7 & 44.3 \\ 
\enddata        
\end{deluxetable*}

Power spectra are computed following the procedure outlined in Section~\ref{sec:powspec} separately for each region (south and equator) 
and for each season pair. The entire set of spectra along with their covariance is  passed on to the likelihood code that forms the basis of 
parameter constraints. Although combined spectra are not used in the actual analysis, in this section we discuss various combinations of power spectra
for  purposes of comparison and systematic tests. Note that \actS\ and \actE\ spectra cannot be trivially combined as residual Galactic cirrus 
contribution to the two regions are different.  Therefore, we subtract the best fit residual cirrus model (as discussed in \citealt{dunkley/etal:prep})
from the estimated power spectra before combining  \actE\ and \actS\ spectra. To simplify the presentation, all figures in this section portray 
 dust-subtracted spectra.  Another complication arises due to the different geometries and masking pattern of the \actE\ and \actS\ maps, which cause  
  the theoretical  bandpowers for these regions to be in principle different, although the actual differences are small. 
  Also, due to subtle variations of the beam profile from one season to another, the beam uncertainties in individual season spectra are  slightly different.
   All these subtleties prompted the separate treatment of power 
spectra in the likelihood.  In this section, we neglect these subtleties and  combine spectra, with inverse variance weight,  across season pairs and regions of the sky. We warn the reader hat such combinations are for visualization purposes only. 
Fig.~\ref{fig:equAndSouth} shows the \actE\ and \actS\ spectra  combined across the different observing seasons, along with their corresponding theoretical band powers.  The power spectrum  combined across all  seasons and across the \actE\ and \actS\ strips  is displayed in Fig.~\ref{fig:uberSpec}.  The corresponding  band power values and uncertainties are tabulated in Table~\ref{tab:spec_table}.  
These plots portray how our pipeline is able to produce an estimate of the power spectrum over the entire multipole range of $500-10000$.  Over the multipole 
range of $\ell \simeq 500-2500$ these spectra clearly show the Silk damping tail of the CMB power spectrum, 
while on smaller angular scales ($\ell \sim 2500-10000$) a clear excess from  the frequency-dependent Sunyaev Zel'dovich effects and extragalactic foregrounds  (radio and infrared point sources) is clearly visible (these contributions are further discussed in \citealt{dunkley/etal:prep}). 

 Finally, we  display, in Fig.~\ref{fig:stateOfArt}, the state of the art in CMB temperature power spectrum 
measurements down to the damping tail  where we plot the WMAP 7-year spectrum, the  inverse variance combined \actE+\actS\ spectrum, and the 
recent SPT power spectrum \citep{story/etal:prep}. 

\subsection{Power spectrum with alternative binning}
\label{ssec:altBins}  
As discussed in Section~\ref{ssec:binning}, the choice of large bins for our main power spectrum result was motivated by the need for 
keeping the bandpowers minimally correlated. It is of interest, however, to ask how the spectrum would have looked with smaller bins 
of width $\Delta \ell = 50$ over the damping tail,  as was done in \citetalias{das/etal:2011}.  Such a result is shown in Fig.~\ref{fig:altBins}. 
Note that the first through the eighth peak of the CMB can be clearly seen with this binning. We do not pursue this binning  any further 
for the aforementioned reasons. 

  \subsection{Derived CMB-only power spectrum}
\begin{figure*}[htbp]   
\begin{center}
\includegraphics[width=1.\columnwidth]{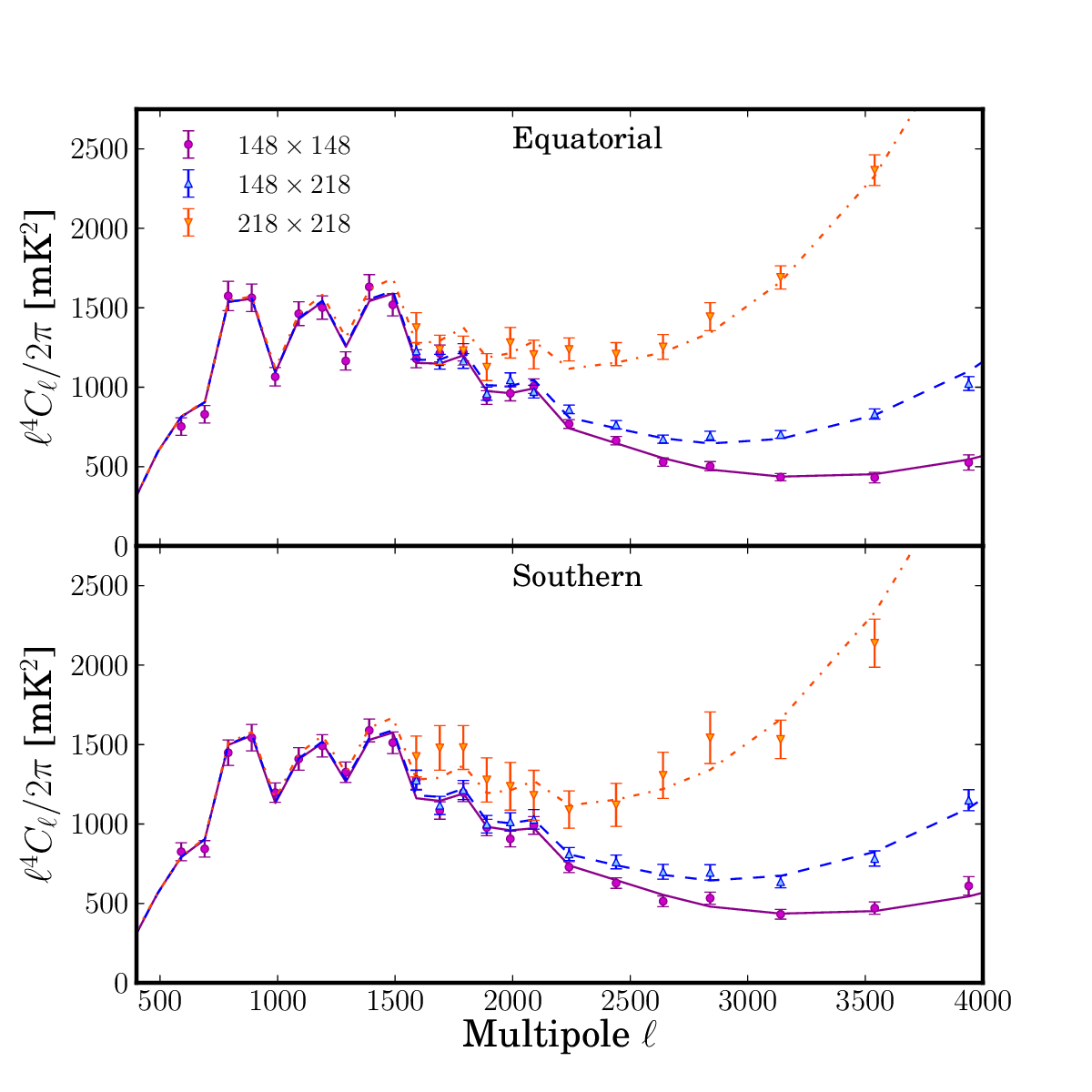}\includegraphics[width=1.\columnwidth]{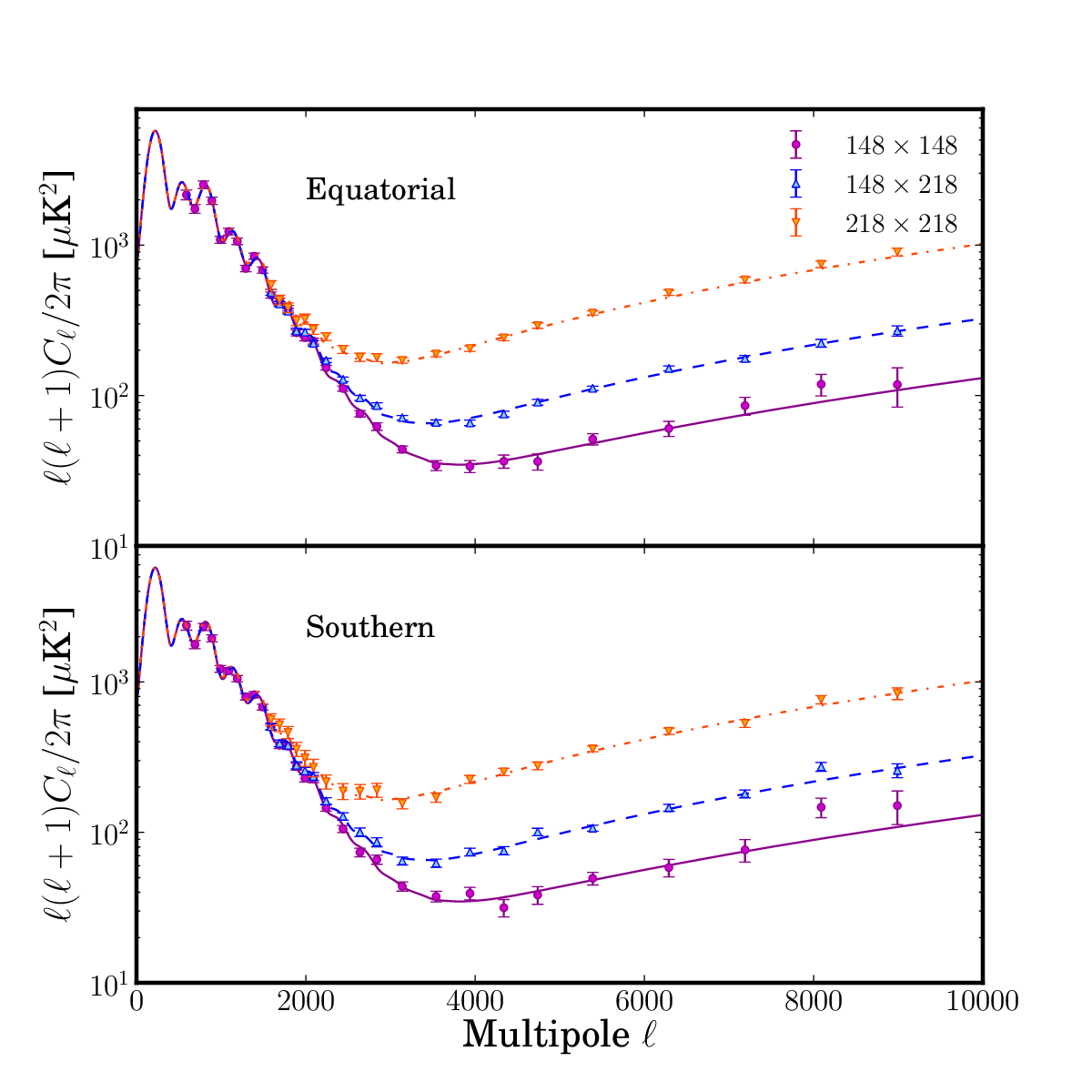}
\caption{Combined multi-season power spectra for the \actE\ Strip (upper panels) and the \actS\ Strip 
(lower panels). For the \actE\ and \actS\ data points,   the  corresponding best fit residual cirrus model (as discussed in \citealt{dunkley/etal:prep}) has been subtracted. The left hand panel shows a linear scale zoomed-in version of the spectrum  with an $\ell^4$ scaling to emphasize the 
higher order acoustic features. The lines show the binned version of the best fit model for each frequency pair including CMB secondaries  and foregrounds from our companion paper \citet{dunkley/etal:prep}.  The right panel shows the entire range of the computed 
spectrum on a log-linear scale with the conventional $\ell (\ell + 1) $ scaling. The lines show the unbinned version of the best fit model from \citet{dunkley/etal:prep}.\label{fig:equAndSouth}}
\label{blah}
\end{center}
\end{figure*}

\begin{figure*}[htbp]
\begin{center}
\includegraphics[width=1.5\columnwidth]{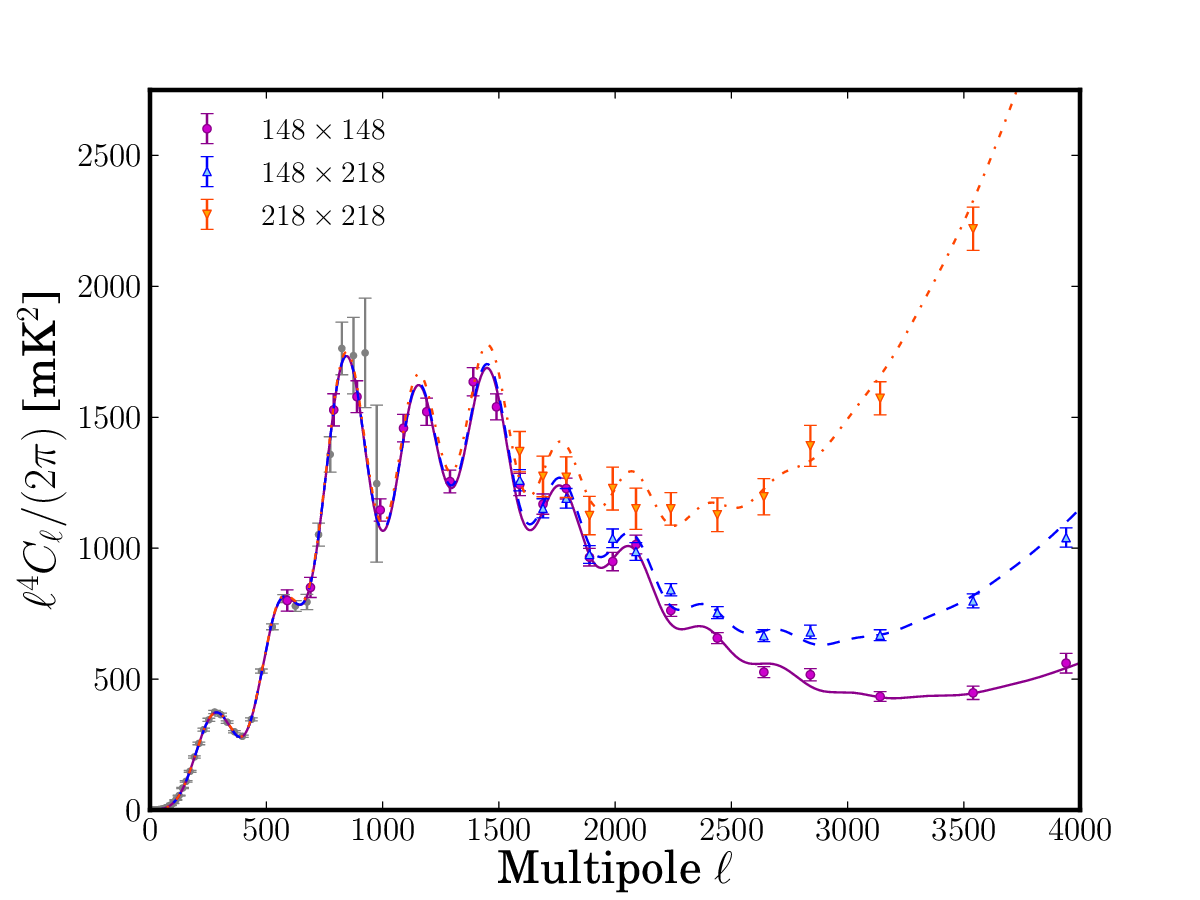}
\includegraphics[width=1.5\columnwidth]{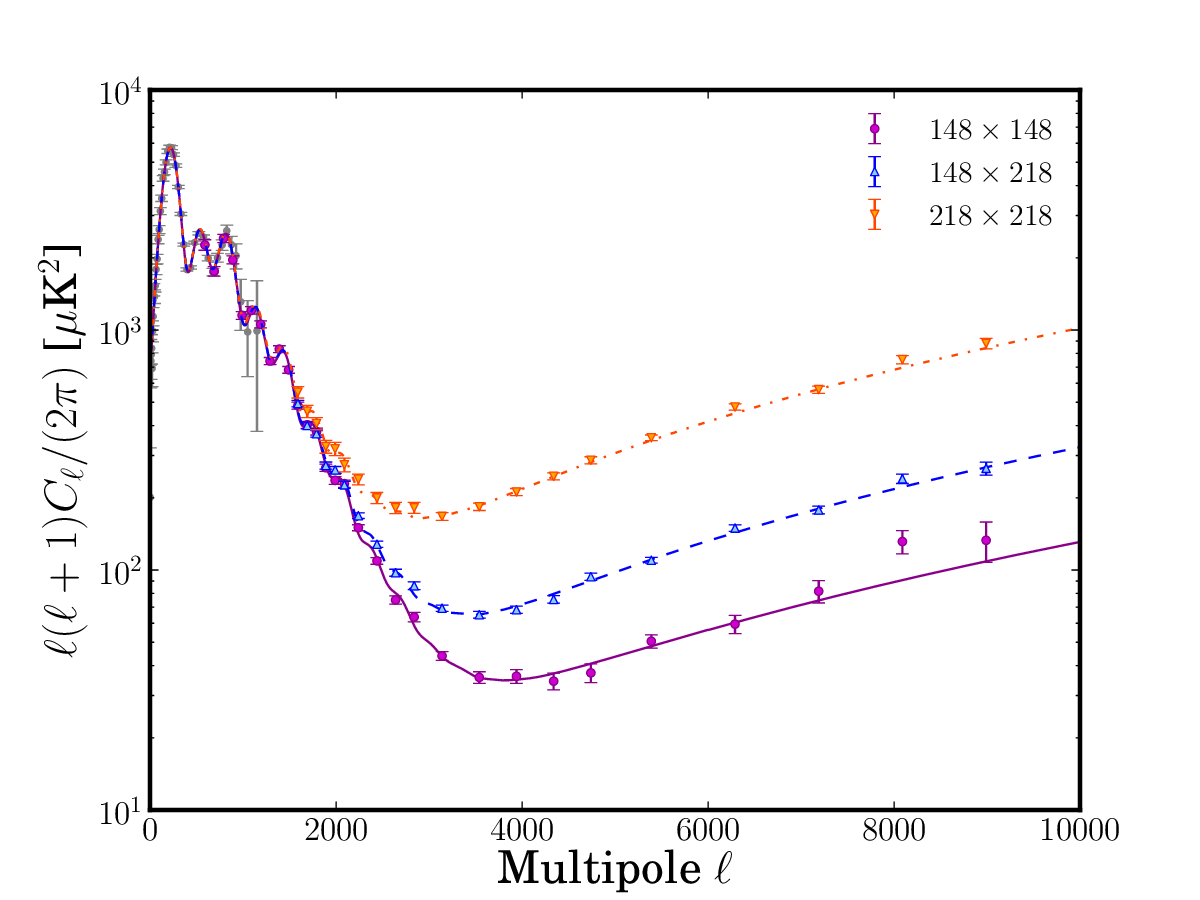}
\caption{Multifrequency power spectra combined across all  seasons and  the  \actE\ and \actS\ regions. Before combining the \actE\ and \actS\ data points,   the  corresponding best fit residual cirrus model (as discussed in \citealt{dunkley/etal:prep}) has been subtracted. The upper panel shows the  $\ell = 500-4000$ portion of 
the power spectrum on a linear scale with an $\ell^4$ scaling to emphasize the higher order acoustic peaks. The lower panel shows the entire 
range of the computed spectra  with the $\ell(\ell+1) $ scaling.  The lines in either case show the best fit models for each frequency pair including CMB secondaries
and foregrounds from our companion paper \citet{dunkley/etal:prep}. The grey data points represent  the power spectrum from the WMAP seven-year data 
release  \citep{larson/etal:2011}.  \label{fig:uberSpec}}
\label{blah}
\end{center}
\end{figure*}

\begin{figure*}[htbp]
\begin{center}
\includegraphics[width=2\columnwidth]{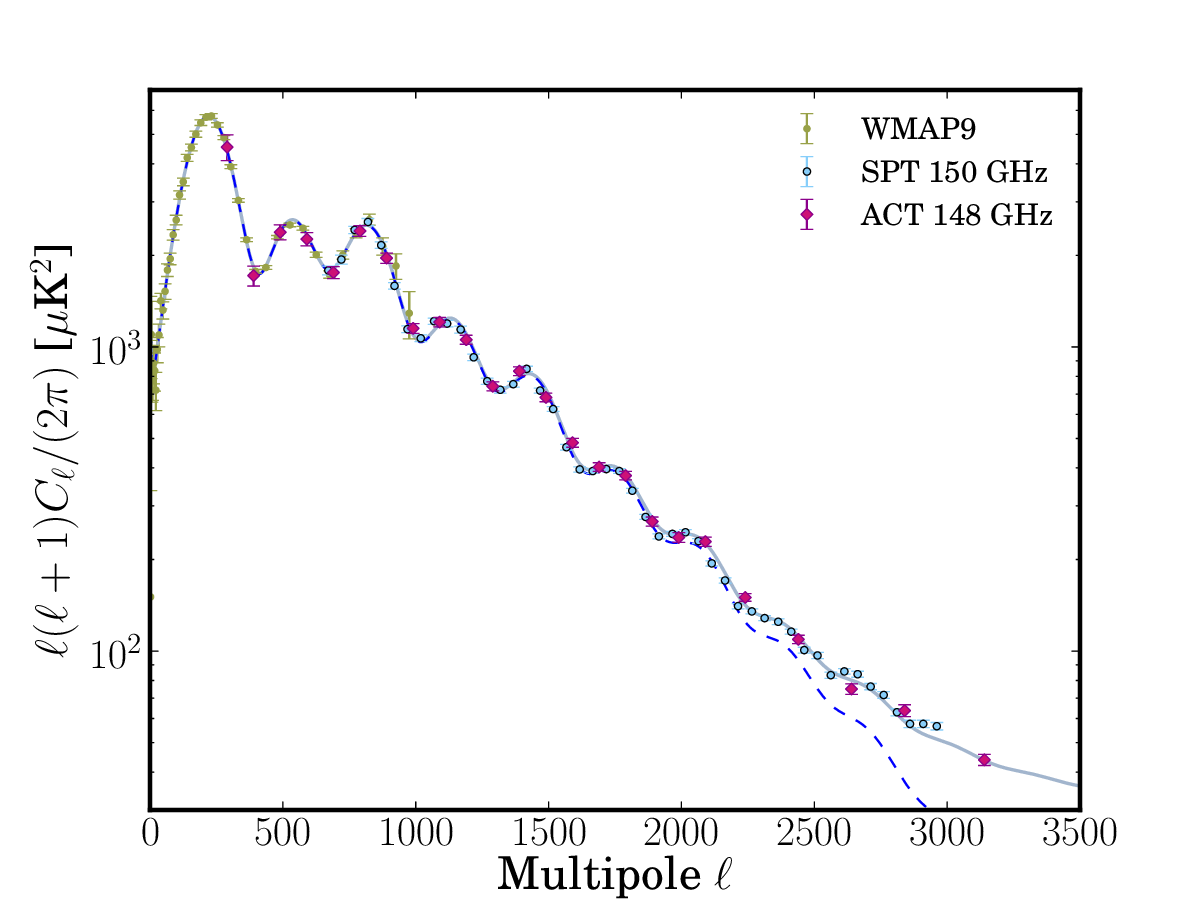}
\caption{State of the art of CMB temperature power spectrum measurements from the WMAP 9-year data release \citep{bennett/etal:prep, hinshaw/etal:prep}, 
the South Pole Telescope \citep{story/etal:prep} and ACT  (this work). The solid line shows the best fit model to the ACT \arone\ data combined with 
WMAP 7-year data \citep{larson/etal:2011}. The dashed line shows the CMB-only component of the same best fit model. Although we compute the power spectrum down to $\ell=200$, we do not use data below $\ell=540$ in the analysis.\label{fig:stateOfArt}}
\end{center} 
\end{figure*}

\begin{figure*}[htbp]
\begin{center}
\includegraphics[width=2\columnwidth]{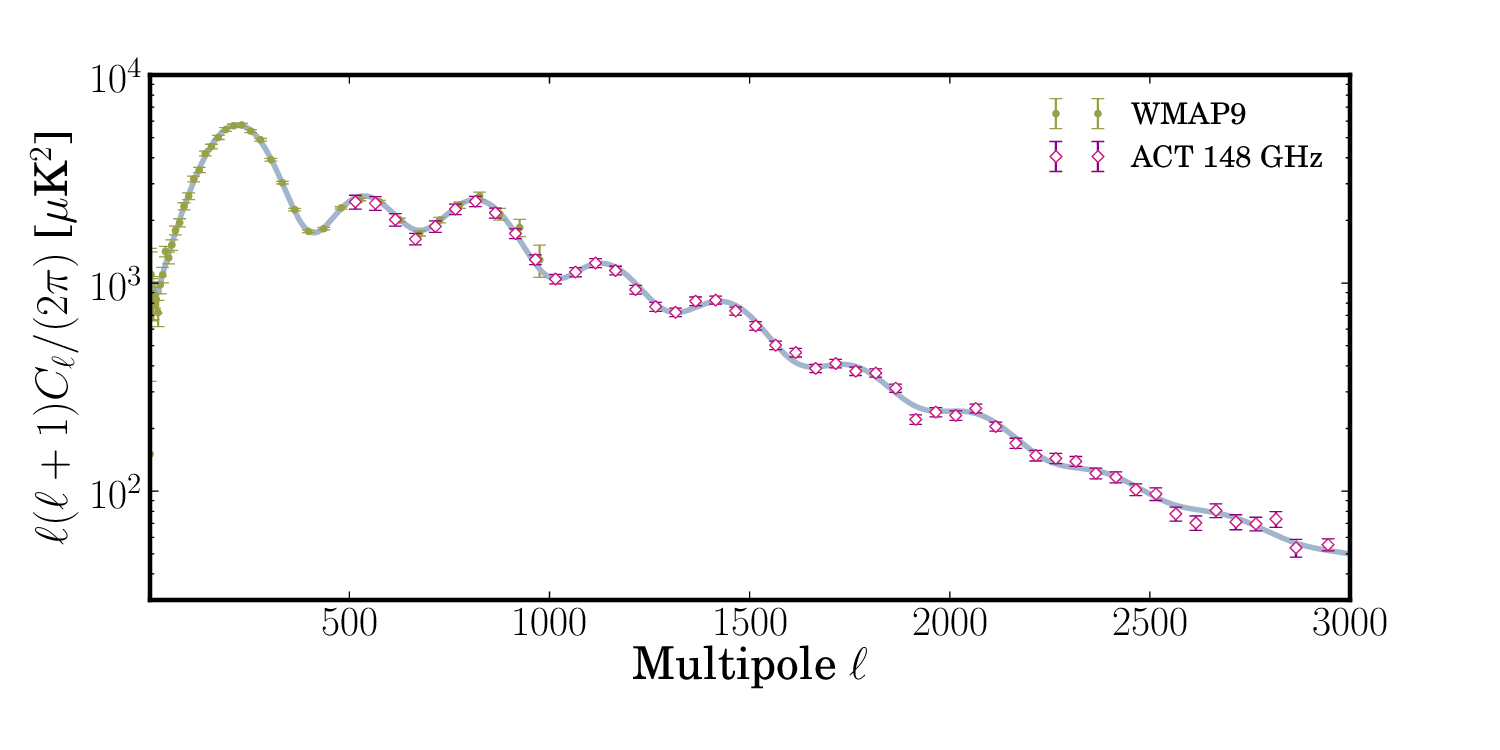}
\caption{Combined \actE\ + \actS\ \arone\  power spectrum computed with alternate binning shown alongside the WMAP 9-year data \citep{bennett/etal:prep, hinshaw/etal:prep}.
Note  that with these smaller  bins, the contours of the first seven acoustic peaks of the CMB power spectrum can be clearly seen.  The bandpowers are significantly correlated
at this bin size, and a precise estimate of the bin to bin correlation is computationally prohibitively costly.  The solid line shows the best fit model to the ACT \arone\ data combined with  WMAP 7-year data \citep{larson/etal:2011}.\label{fig:altBins}}
\end{center} 
\end{figure*}

\begin{figure*}[htbp]
\begin{center}
\includegraphics[width=2\columnwidth]{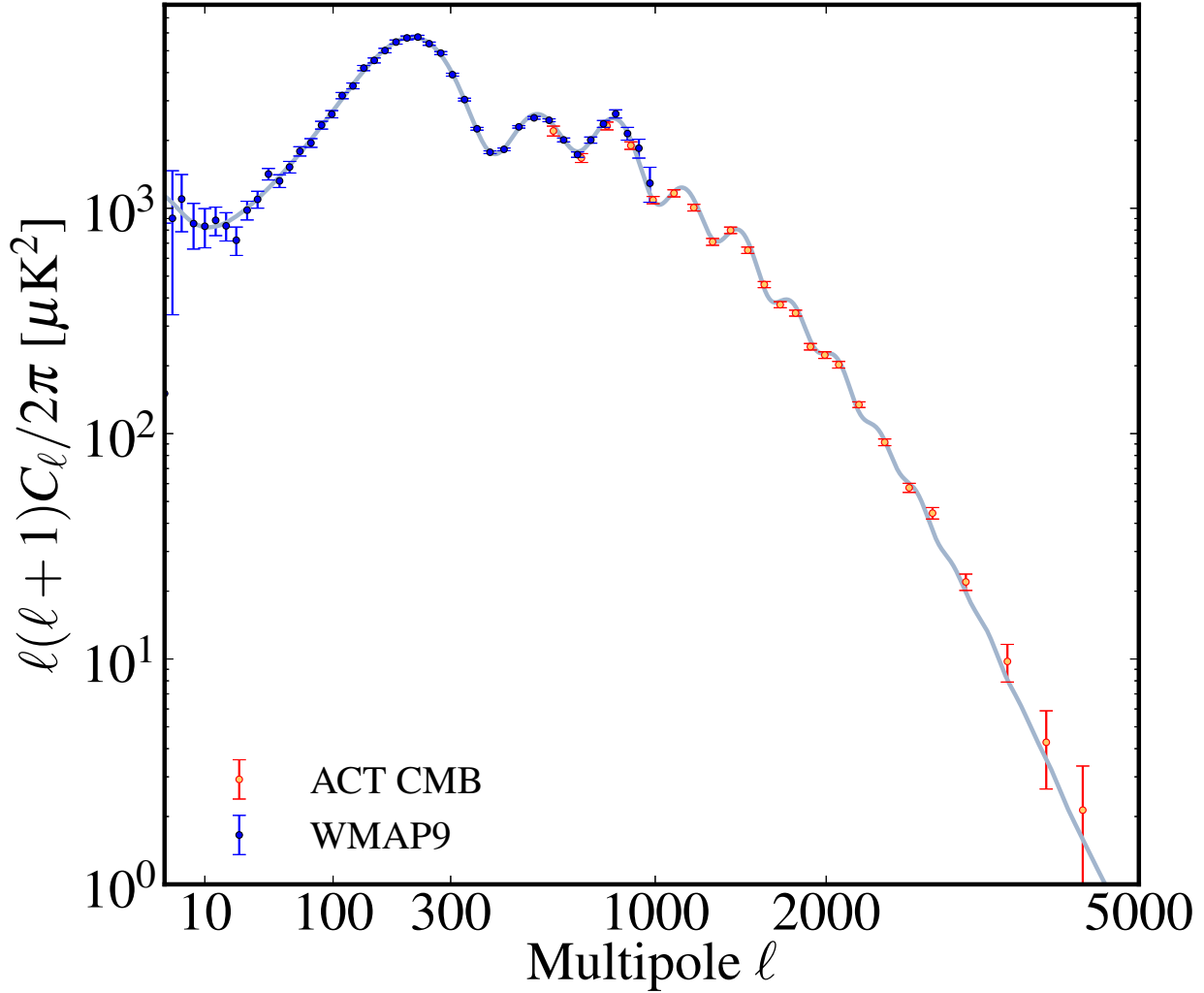}
\caption{The CMB power spectrum estimated from ACT, shown with the spectrum from the WMAP 9-year data \citep{bennett/etal:prep,hinshaw/etal:prep}. The errors include uncertainty due to foreground and SZ emission, as well as the relative calibration of the 148 and 218 GHz channels, and beam uncertainty. The full covariance matrix is derived in \citet{dunkley/etal:prep}. The solid line shows the CMB-only component of the best fit model for the ACT data combined with the WMAP 7-year data.}
\label{fig:derivedCMB}
\end{center}
\end{figure*}

 The \actE\  and \actS\ spectra shown in Fig.~\ref{fig:equAndSouth} include the primary CMB signal as well as power from foregrounds and SZ. We show the estimated primary CMB spectrum from ACT in Fig.~\ref{fig:derivedCMB},  derived in \citet{dunkley/etal:prep}. There, the multi-frequency spectra are used to estimate the CMB in bandpowers for ACT-E and ACT-S, simultaneously with the SZ and foregrounds components. The CMB spectra for ACT-E and ACT-S are then coadded for display. No assumptions are made about the cosmological model, only that the CMB is blackbody. Using the multi-frequency data to separate components, the CMB power can be recovered out to multipoles of $\ell\sim 3500$.

\subsection{Systematic Tests}
In order to check for systematics in the map-making and  power spectrum estimation pipelines, we perform various tests on the data.
These are constructed such that the sky signal cancels between the various splits of the data, and only systematic effects remain. 
We  test that the power spectrum obtained is the same in each season, in all time splits, from different parts of the array, with and without data near the telescope  turnaround points, from different directions in Fourier space, and for different regions of the sky. 
    The statistics from a subset of these  tests are summarized in Tables~\ref{cross_null_test_table}, \ref{TOD_null_test_table},  \ref{detector_in_out}, and \ref{detector_turnaround}. Given that the spectra are computed individually and then included in the likelihood with the full covariance of the different frequencies and seasons, we compute the null tests on  each subset of  data, both for \actE\ and \actS, and for different seasons.
\subsubsection{Cross season nulls}
\begin{figure}[htbp!]
\begin{center}
$\begin{array}{@{\hspace{-0.25in}}l}
\includegraphics[width=0.54\textwidth]{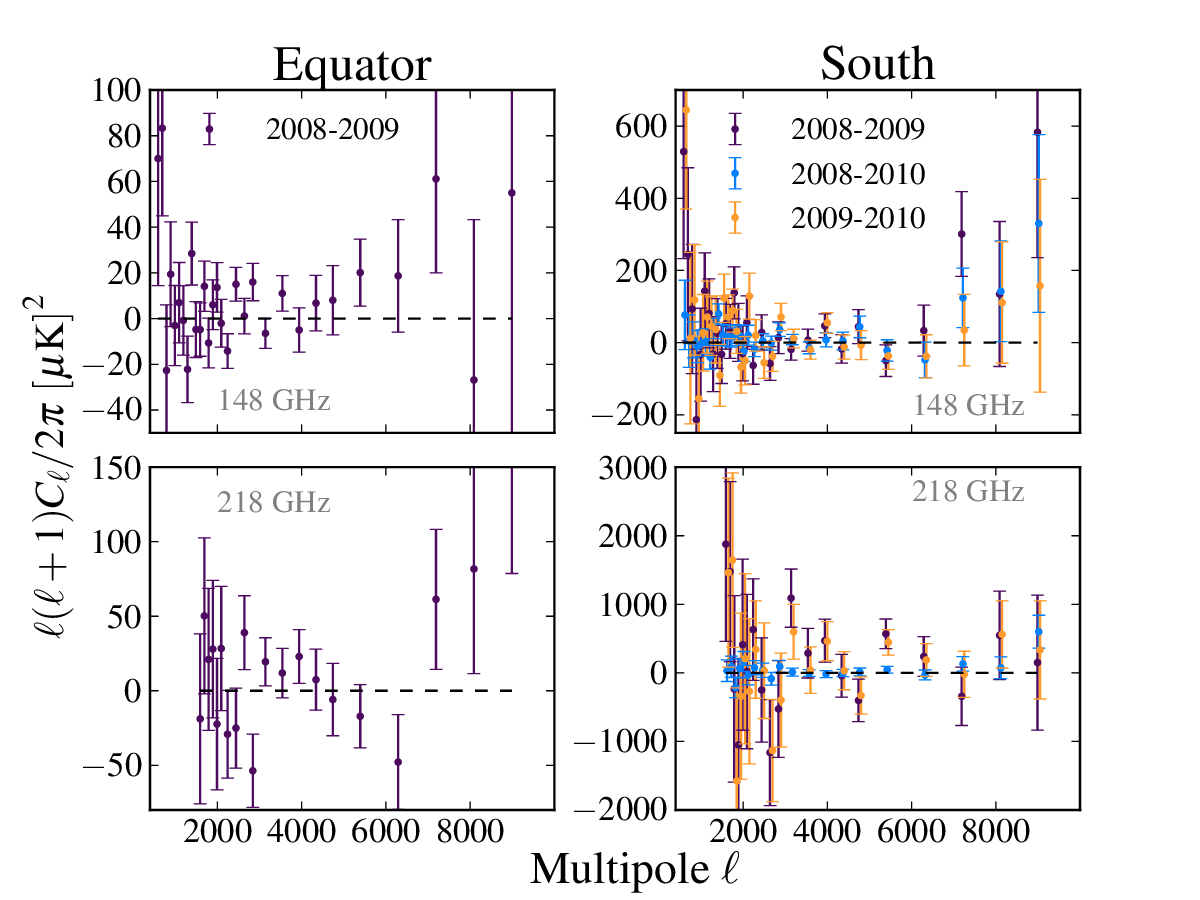}  \\
%,trim = 0mm 40mm 0mm 60mm, clip
 \end{array}$
\caption{Cross season null test for the ACT data. The top row
illustrate the \arone\ cross season null tests for \actE\ (top left)
and \actS\ (top right), while the bottom row show the \artwo\ cross
season nulls.The $\chi^2$ values for the fit are presented in
Table~\ref{cross_null_test_table} and Figure~\ref{fig:allchi2}.
\label{fig:cross}}
\end{center}
 \end{figure}

\begin{table}[htbp!]
\caption{\small{Null test $\chi^2 $ values for the season consistency tests performed on the ACT data. The probability to exceed (PTE) the 
$\chi^2$ is shown in  parentheses.}\label{cross_null_test_table}}
\begin{center}
\hspace{-0.3in}
\begin{tabular}{|c|c|ccc|c|}
\hline
Frequency&Region& Seasons & Seasons  & Seasons & dof\\
\hline
\hline
 &&2008-2009&2009-2010&2008-2010& \\
\hline
148 GHz&South&32.2 (0.36) &30.7 (0.43) &35.0 (0.24) &30\\
\hline
&Equator&- &39.7 (0.11) &-&30\\
\hline
\hline
220 GHz&South&27.7 (0.11) &15.9 (0.72) &21.5 (0.37) &20\\
\hline
&Equator&- &24.2 (0.23) &-&20\\
\hline
\end{tabular}
\end{center}
\end{table}
First, we test the year-to-year consistency of power spectra.
In order to account for differences in the ACT beam from one observing season to the next, we convolve the map from one season with 
the beam profile of the other season being differenced, so  that each map effectively has the same beam transfer 
function. Then we difference the corresponding splits from  seasons $s_1$ and $s_2$:
\begin{equation}
\Delta T^{i}(\nhat)\equiv [T^i_{s_1}(\nhat) -T^i_{s_2}(\nhat)],
\label{eq:cross_diffs}
\end{equation}
where i = 1,2,3,4 represent the split index. 
The pixel weight map $W$ corresponding to these difference splits is computed as:
\begin{equation}
W^{-1} = W^{-1}_{s_1} + W^{-1}_{s_2},
\end{equation}
where $W_{s_1}$  is the  total $N_{\rm{obs}}$  map for season $s_1$ etc. 
As with the other null tests, the azimuthal weighting is computed using the weights from the full data spectrum run. 
Figure~\ref{fig:cross} shows these test for the ACT data, while the individual $\chi^2$ values for the tests are summarized in Table~\ref{cross_null_test_table}. We find all spectra computed in this way to be consistent with null.

\subsubsection{Split Nulls}

\begin{table*}[htbp!]
\caption{\small{$\chi^2$ (PTE)  values for the TOD split null tests performed on the ACT data.}\label{TOD_null_test_table}}
\begin{center}
\begin{tabular}{|c|c|c|ccc|c|}
\hline
Frequency&Region& Season& &TOD&&dof\\
\hline
&&&  (1-2)x(3-4)&(1-3)x(2-4)&(1-4)x(2-3)&\\
\hline
148 GHz & South& 2008 &19.7 (0.92) &37.7 (0.16) &37.1 (0.17) &30\\
& &2009 &30.3 (0.45) &31.8 (0.38) &22.6 (0.83) &30\\
& &2010 &35.7 (0.22) &28.6 (0.54) &21.5 (0.87) &30\\
\hline
&  Equator&2009 &33.9 (0.29) &26.5 (0.65) &40.9 (0.09) &30\\
& &2010 &34.6 (0.26) &35.6 (0.22) &24.0 (0.77) &30\\
\hline
\hline
220 GHz & South& 2008 &33.1 (0.03) &28.2 (0.10) &15.3 (0.76) &20\\
& &2009 &14.4 (0.81) &11.3 (0.94) &14.8 (0.79) &20\\
& &2010 &8.8 (0.99) &16.0 (0.72) &21.0 (0.40) &20\\
\hline
&  Equator&2009 &24.9 (0.21) &19.3 (0.50) &13.3 (0.87) &20\\
& &2010 &11.8 (0.92) &16.3 (0.70) &14.0 (0.83)&20\\
\hline
\end{tabular}
\end{center}
\end{table*}

\begin{figure}[htbp!]
\begin{center}
$\begin{array}{@{\hspace{-0.25in}}l}
\includegraphics[width=0.54\textwidth]{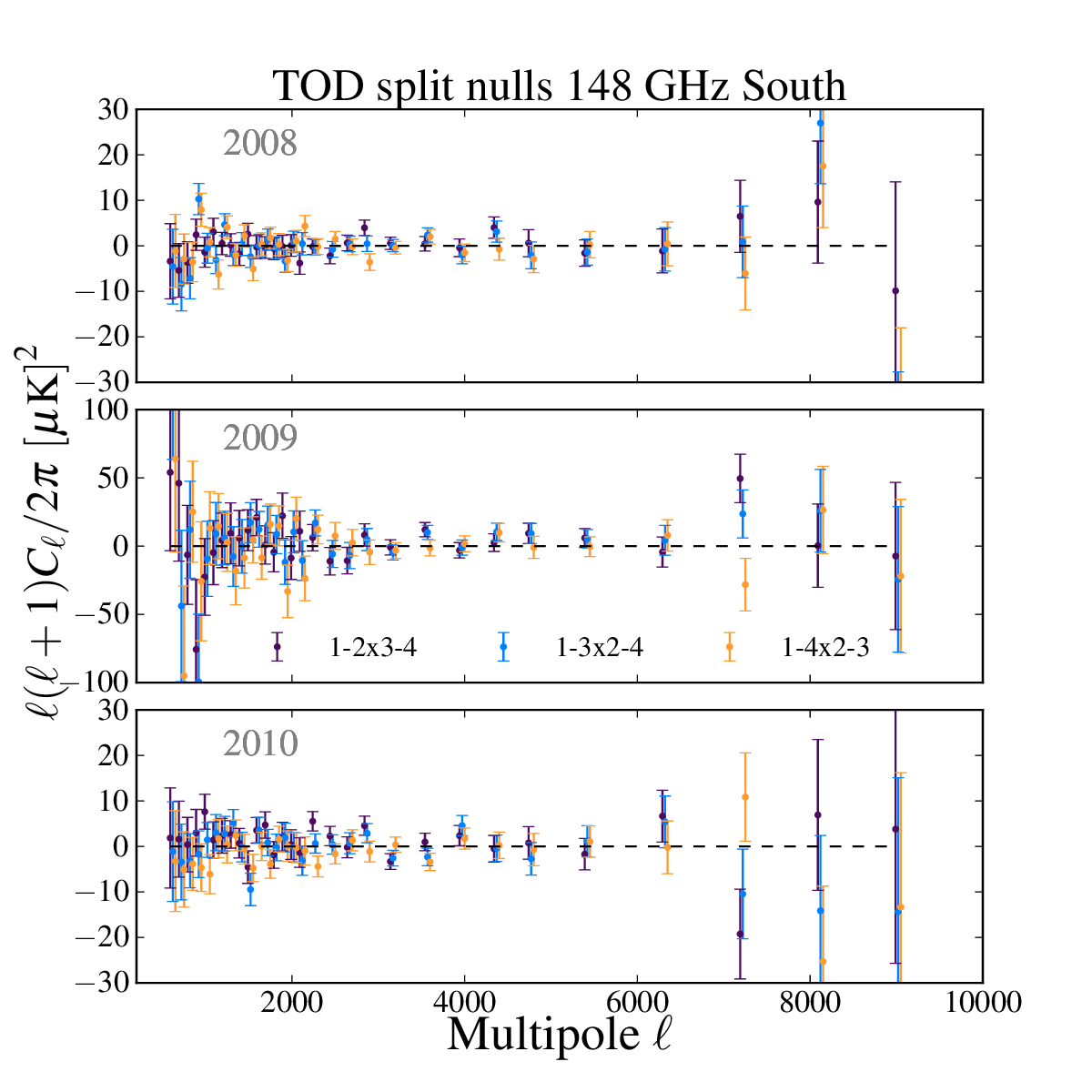}%,trim = 0mm 40mm 0mm 60mm, clip
 \end{array}$
\caption{TOD null test for the 148 GHz Southern strip, from 2008 (top panel) to 2010 (bottom panel). For each year, three TOD nulls are created from the combinations described in Eq.~\ref{eq:tod_diffs}. The $\chi^2$ values for the null test are summarised in Table~\ref{TOD_null_test_table}.\label{fig:tod}}
\end{center}
 \end{figure}
As discussed in Section~\ref{sec:powspec}, the data in each season are separated into four splits in such a way that the detector 
noise is  independent from one split to another. Therefore, the difference between any two splits  should be consistent with noise 
and the signal should subtract away. We test this  by generating difference maps from each pair, and computing the two-way cross spectra from independent pairs of difference maps, e.g.:
\begin{eqnarray}
\label{eq:tod_diffs}
T^{12}(\nhat)\equiv[T^1(\nhat) -T^2(\nhat)]/2 \nonumber \\
T^{34}(\nhat)\equiv[T^3(\nhat) -T^4(\nhat)]/2.
\end{eqnarray}
The difference maps are expected to contain noise but no
residual signal. We estimate the cross-spectrum of the
difference maps, $\hat{C}_b =\avg{ T^{12}T^{34}}, $ and the other two permutations of the differences ($\avg{ T^{13}T^{24}}$ and $\avg{ T^{14}T^{23}}$). 
These difference maps are downweighted by the same weight maps used to construct the full power spectrum. Similarly, the azimuthal weights are borrowed  from the full data spectrum run.  The three difference spectra are shown in Figure~\ref{fig:tod} for the 148 GHz \actS\ data set. The statistics corresponding to this test are shown in Table~\ref{TOD_null_test_table}. The spectra are found to be consistent with a null signal, as expected. 

\subsubsection{In/Out nulls}
In order to test for systematic detector asymmetries, we make a map  using data from  detectors  from the inner region of the array,  and another map from detectors along the edges, and compute the differences between the two maps:
\begin{eqnarray}
\nonumber T^{12}_{io}(\nhat) &\equiv& [T^1_{o}(\nhat) -T^2_i(\nhat)]/2 \nonumber \\
 T^{34}_{io}(\nhat) &\equiv& [T^3_o(\nhat) -T^4_i(\nhat)]/2
\label{eq:io_diffs}
\end{eqnarray}
where the $i$ and $o$ label the inner and outer parts of the detector array respectively. The full set of  $\chi^2$ values are summarized in Table~\ref{detector_in_out}. In general we see no trend for differences as a function of detector position; the null tests are consistent with no signal.
\begin{table*}[htbp!]
\caption{\small{Null test $\chi^2$ (PTE) values for the inner vs outer detectors.}\label{detector_in_out}}
\begin{center}
\begin{tabular}{|c|c|c|ccc|c|}
\hline
Frequency&Region& Season& & In/Out &&dof\\ \hline
& &&(1o-2i)x(3o-4i)&(1o-3i)x(2o-4i)&(1o-4i)x(2o-3i)&\\
\hline
148 GHz & South&2008&26.2 (0.66) &28.2 (0.56) &26.2 (0.67) &30\\
& &2009 &37.7 (0.16) &17.3 (0.97) &27.3 (0.61) &30\\
& &2010 &34.7 (0.26) &27.3 (0.61) &25.6 (0.70) &30\\
\hline
&  Equator&2009 &32.0 (0.36) &24.8 (0.74) &26.2 (0.66) &30\\
& &2010 &32.2 (0.36) &36.3 (0.20)& 38.6 (0.13)  &30\\
\hline
\hline
218 GHz & South&2008&25.3 (0.19) &23.9 (0.24) &21.0 (0.40) &20\\
& &2009 &13.7 (0.85) &12.3 (0.91) &25.8 (0.17) &20\\
& &2010 &13.8 (0.84) &15.0 (0.78) &26.3 (0.16) &20\\
\hline
&  Equator&2009 &23.4 (0.27) &22.6 (0.31) &27.9 (0.11) &20\\
& &2010 &9.9 (0.97) &25.3 (0.19) &9.2 (0.98) &20\\
\hline
\hline
\end{tabular}
\end{center}
\end{table*}

 \begin{figure}[htbp!]
\begin{center}
$\begin{array}{@{\hspace{-0.05in}}l}
\includegraphics[width=0.5\textwidth]{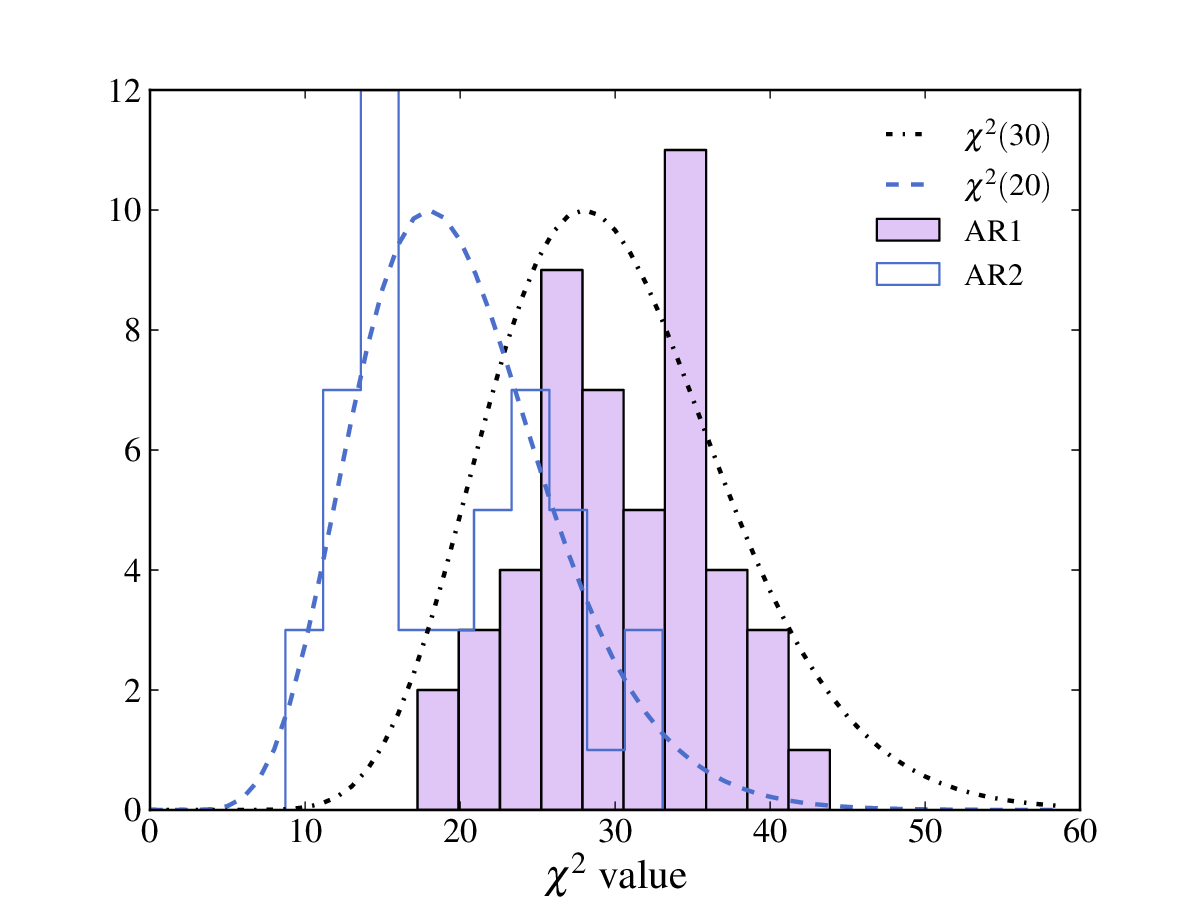}%,trim = 0mm 40mm 0mm 60mm, clip
 \end{array}$
\caption{The reduced $\chi^2$ values for all null tests. The blue histogram is computed for the $\chi^2$ values from the 218 GHz null tests, while the purple histogram shows the same null tests for the 148 GHz maps. The black dashed and dot-dashed lines show the theoretical distributions for 20 (AR2) and 30 (AR1) degrees of freedom respectively, normalized to match the frequency of the histograms. The $\chi^2$ values presented here are given in Tables~\ref{cross_null_test_table}, \ref{TOD_null_test_table}, \ref{detector_in_out} and \ref{detector_turnaround}. \label{fig:allchi2}}
\end{center}
 \end{figure}

\subsubsection{Turnarounds}
Another null test, based on cutting out data around telescope turnaround 
probes the consistency of data taken
with the telescope accelerating as it reverses direction at
the ends of the scan (turnarounds). In the maps used for the power spectrum estimation, the data taken during the turnaround is included.
We test for any artifacts generated by the acceleration at turnaround by taking the difference of maps with and without turnaround data.
Maps are made cutting data near the turnarounds,
amounting to removing $\approx 10\%$ of the total data. This loss of data affects the two sky regions differently. In the southern patches, the loss of data is uniform and leads to a slight increase in striping in the maps, whereas in the equatorial patches, removing the turnarounds removes data at the upper and lower edges of the maps. Hence, for these tests we compute the differences using  an equatorial region which is slightly narrower in the declination direction 
($2\fdeg7$ as opposed to $3\deg$ wide). 
Two difference
maps are made by pairing one split of the standard
map with a different split of the new maps (we avoid differencing
the same splits as they have very similar noise
structure), and a two-way cross-power spectrum is produced.
Any artifact due to the turnaround would be
left in these difference maps and might produce excess
power.
We compute the turnaround cuts as a function of season, frequency range and area on the sky. The reduced $\chi^2$ values are summarized in Table~\ref{detector_turnaround}. Again, we find that the difference maps have spectra consistent with no signal.

\subsubsection{$\chi^2$ Distribution}
While the null tests are performed for different subsets of the data, we combine the statistics from the null tests together to test for consistency globally. We restrict the range of the \artwo\ spectrum to be $\ell > 1500$, hence the \artwo\ spectrum contains 20 degrees of freedom, while the \arone\ spectrum containts 30 degrees of freedom. We show the distribution of $\chi^2$ values and the theoretical $\chi^2$ distribution for the two cases  in Figure~\ref{fig:allchi2}. This shows that the null tests are broadly consistent with being drawn from a $\chi^2$ distribution for the number of degrees of freedom.

\subsubsection{Isotropy}

\begin{figure}[htbp]
\begin{center}
\includegraphics[width=\columnwidth]{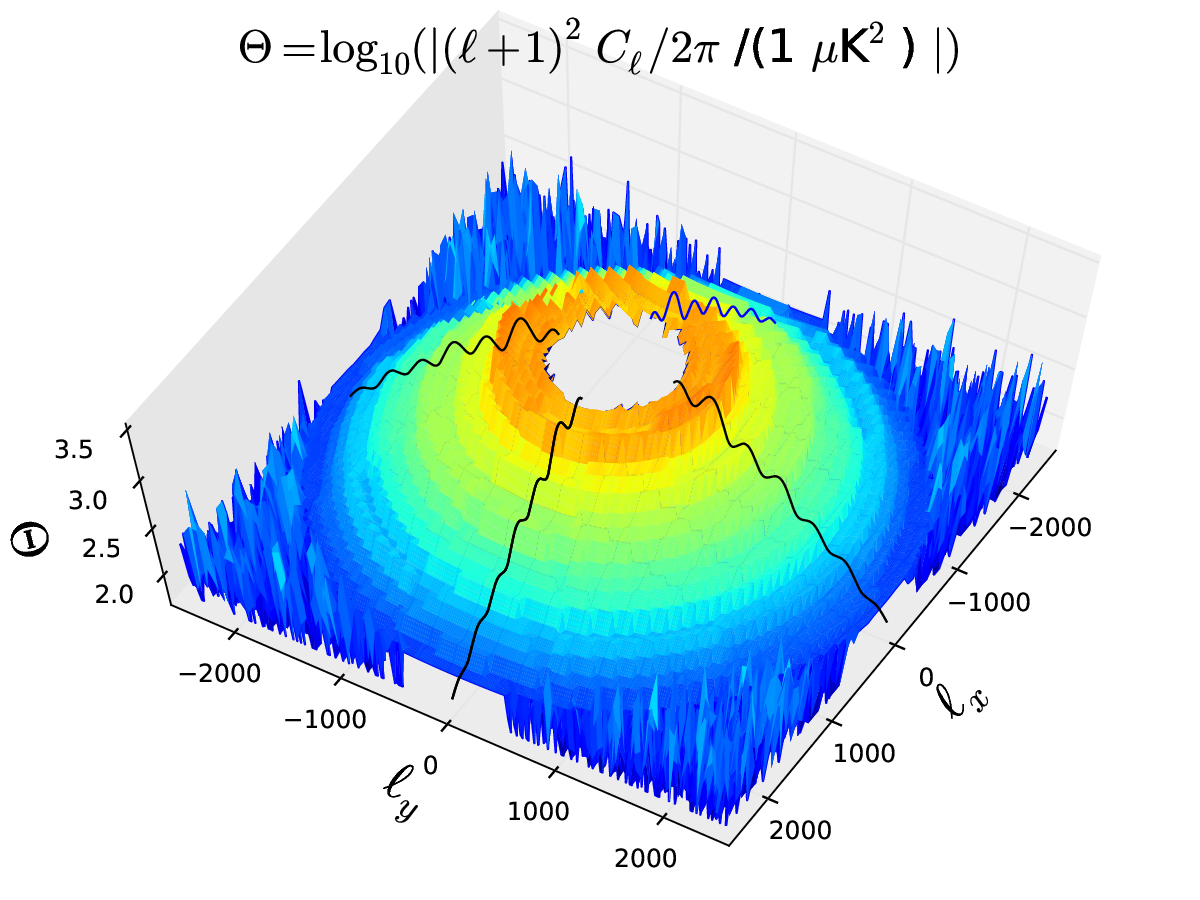}
\caption{The two dimensional  \arone\ cross-power  spectrum co-added across the ACT-E patches  and  seasons. 
For $\ell<2500$  smoothing by a small kernel has been applied.  The acoustic features in the power spectrum are  clearly visible. 
 At $\ell>2500$, where the instrument noise dominates, we display the raw spectrum.   The black lines represent the CMB-only theory and have been plotted to guide the eye.  \label{fig:power2D}}
\end{center}
\end{figure}

\begin{table*}[htpb!]
\caption{\small{Null test $\chi^2$ values for  turnarounds}\label{detector_turnaround}}
\begin{center}
\begin{tabular}{|c|c|c|ccc|c|}
\hline
Frequency&Region& Season& &Turnarounds&&dof\\ 
\hline 
& &&(1t-2nt)x(3t-4nt) &(1t-3nt)x(2t-4nt)&(1t-4nt)x(2t-3nt)&\\
\hline
148 GHz & South&2008&21.9 (0.85) &24.8 (0.74) &34.6 (0.26) &30\\
& &2009 &28.3 (0.56) &35.3 (0.23) &26.7 (0.64) &30\\
& &2010 &30.5 (0.44) &30.4 (0.44) &35.0 (0.24) &30\\
\hline
&  Equator&2009 &27.4 (0.60) &29.3 (0.50) &34.0 (0.28) &30\\
& &2010 &35.1 (0.24) &43.9 (0.05) &21.9 (0.86) &30\\
\hline
\hline
218 GHz & South&2008&25.9 (0.17) &25.3 (0.19) &20.6 (0.42) &20\\
& &2009 &15.8 (0.73) & 13.0 (0.88) &14.3 (0.82) &20\\
& &2010 &11.2 (0.94)&11.6 (0.93)& 21.1 (0.4) &20\\
\hline
&  Equator&2009&31.3 (0.06) &19.4 (0.5)  &16.6 (0.68) & 20\\
& &2010 &14.1 (0.82) &24.3 (023) &16.7 (0.67) &20\\
\hline
\end{tabular}
\end{center}
\end{table*}

We test the isotropy of the power spectrum by estimating the power as a function of phase $\theta = \arctan(\ell_y/\ell_x)$.
We compute the inverse-noise-weighted two-dimensional
 spectrum co-added across patches and seasons for the ACT-E region. We show the mean two-dimensional cross-power pseudo spectrum in Figure~\ref{fig:power2D}. The spectrum is symmetric for $\ell$ to $-\ell$, as it is for any real valued
map. To quantify any anisotropy, the power averaged over all multipoles in the range $200 < \ell < 10000$ is computed in wedges of $\theta = 20^\circ,$ and compared to the mean of the entire annulus. No significant deviation from isotropy is detected using this method.  We find that this  result holds 
for \actS\ and \artwo\ maps.

%\tbd{Null and consistency checks}

\subsection{Consistency of \actE , \actS, and SPT spectra} 
\begin{figure}[htbp]
\begin{center}
\includegraphics[width=\columnwidth]{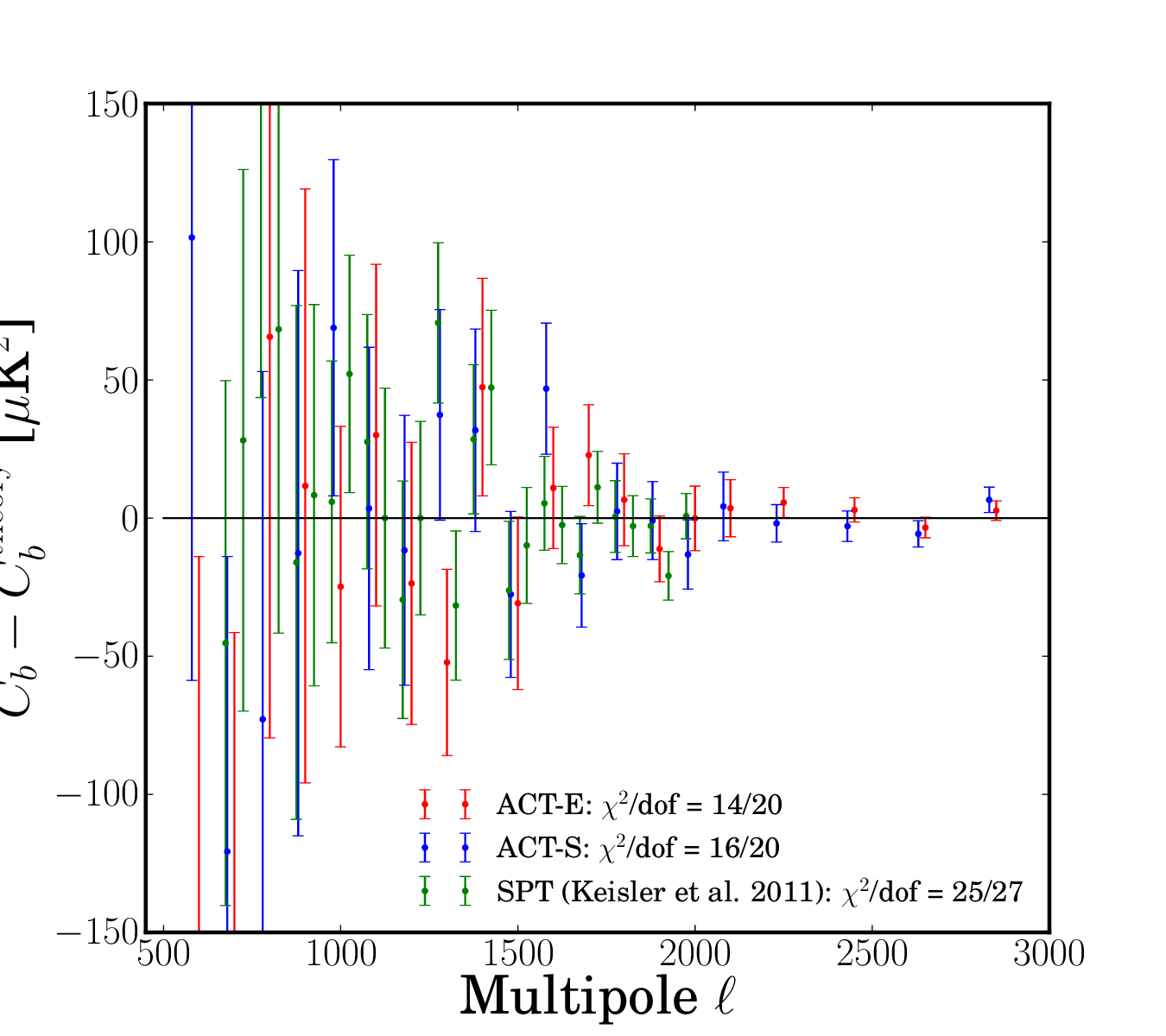}
\caption{The residual power spectra obtained after subtracting the best fit ACT + WMAP model from the \actE, \actS, and SPT power spectrum \citep{keisler/etal:2011}. For the SPT residual the Poisson  point source component of the best fit model is adjusted to reflect the difference 
in point source masking levels between ACT and SPT. The residuals are all null showing the  consistency of these spectra.}
\label{fig:residualSpec}
\end{center}
\end{figure}

As mentioned above, due to the difference in geometry of the equatorial vs. southern patches, the band power binning functions 
for \actE\ and \actS\ are slightly different leading small differences in the binned version of the best fit ACT $+$ WMAP7 model \citep{sievers/etal:prep}.
 Therefore  to test the consistency between \actE\ and \actS\  spectra we check for the nullity of the residuals from their corresponding  binned best fit model. We also consider the consistency of 
the SPT band powers from \citet{keisler/etal:2011}. Care must be taken while computing the residual for the SPT spectrum as the 
point source masking threshold corresponding to that spectrum was different from that of ACT. To correct for this, we adjust the Poisson point 
source component of our best fit model to match the masking level used in \citet{keisler/etal:2011}. The results are shown in Fig.~\ref{fig:residualSpec}, 
 clearly indicating that the three spectra are consistent with  null, and therefore with each other. 

\par
The suite of consistency tests performed here show that our reported spectrum  passes  a wide range of checks for systematic errors in time, detector-space, map-space, and $\ell$-space. 

\section{Gravitational Lensing Analysis }
\label{sec:cmblensing}
\begin{figure}[htbp]
\begin{center}
\includegraphics[width=\columnwidth]{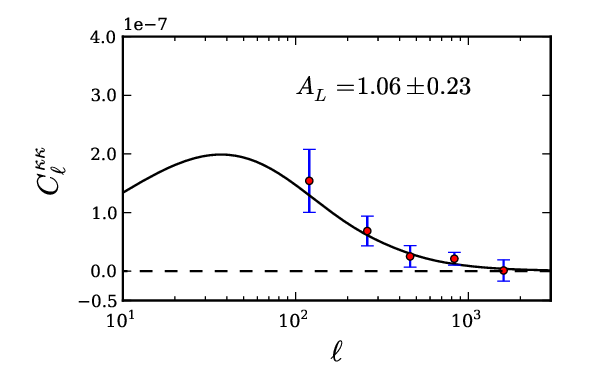}
\caption{CMB convergence power spectrum reconstructed from the \actE\ strip temperature data.    The enhanced 
effective depth of the coadded \actE\ map ($\simeq 18$ $\mu$k-arcmin) compared to its previous version ($\simeq 23$ $\mu$k-arcmin) 
used in \citet{das/etal:2011a} led to an improved detection significance.}

\label{fig:cmbLens}
\end{center}
\end{figure}

Large-scale structure gravitationally deflects the CMB radiation as it passes through the universe, thereby defining a lensing deflection field $\mathbf{d}$ that remaps the observed CMB temperature sky: $T(\nhat) = T_{\rm{unlensed}}(\nhat + \mathbf{d})$. Lensing distorts the small-scale CMB anisotropies, thus modifying their statistical properties, with the lensing deflection locally breaking statistical isotropy and correlating formerly independent temperature Fourier modes (or more intuitively, correlating the CMB temperature with its gradient). The lensed small-scale CMB thus contains not only information about the universe at the last-scattering surface ($z\simeq 1100$), but also encodes information about the cosmic mass distribution at later times and lower redshifts. Using an estimator quadratic in temperature that measures the lensing-induced change in the statistics of the CMB fluctuations from the correlation of formerly independent Fourier modes, one can construct a noisy estimate of the CMB lensing convergence, $\kappa = -\frac12\nabla \cdot \mathbf{d}$, which is a measure of the projected matter density out to high redshifts. The power spectrum of the CMB lensing convergence can be simply obtained from this estimated lensing convergence field, though biases arising from instrument and cosmic-variance-induced noise as well as from higher order corrections to the estimator must be subtracted.  Equivalently the estimation of the lensing power spectrum can be regarded as a measurement of a lensing-induced non-Gaussianity in the CMB temperature four-point correlation function.  Measurements of the lensing power spectrum can place strong constraints on the properties of dark energy and neutrinos, and also serve as a valuable test of the $\Lambda$CDM prediction for structure growth and geometry at redshifts $z\simeq 1-3$.

CMB lensing science has made great advances in recent years. Signatures of CMB lensing were first observed in correlations of the CMB with large-scale structure \citep{smith/etal:2007, hirata/etal:2008}. The lensing power spectrum was first detected at $4\sigma$ significance by the ACT collaboration (\citealt{das/etal:2011a}). The first evidence for dark energy from the CMB alone was also obtained from this measurement by \citealt{sherwin/etal:2011}. A more significant detection (at $6.3~\sigma$) of the lensing power spectrum and more precise cosmological constraints were reported by the SPT collaboration the following year \citep{van/etal:2012}.

 In this work we present a measurement of the lensing power spectrum from the improved ACT maps on the \actE\ strip. Our new measurement of lensing essentially uses the same methodology as \citet{das/etal:2011a}, hereafter \citetalias{das/etal:2011a}. Lensing is measured using a quadratic estimator in temperature; the power spectrum of the CMB lensing convergence is thus a temperature four-point function with the filtering and normalization obtained as in \citetalias{das/etal:2011a}. The bias is calculated in three steps  to make the calculation as robust and model-independent as possible. First, we simulate the bias that is present in the absence of any mode-correlations, removing existing correlations by randomizing the phases of each Fourier mode of the temperature field \citepalias{das/etal:2011a}. This is mathematically equivalent to estimating the so-called $N(0)$ bias from the pseudo-$\cl$ power spectrum measured from this map. The use of measurements rather than simulations makes the bias calculation more robust. Next, we simulate additional small biases from finite-map effects, and anisotropic and inhomogeneous noise by performing lensing power spectrum estimation on unlensed simulated CMB maps with realistic  noise characteristics. After accounting for these two sources of bias, the input lensing power spectrum is recovered fairly accurately in simulations, but there is a small difference between recovered and input power spectra due to higher order corrections. We adopt the small, simulated difference between the recovered and true power spectra as an additional bias and subtract it from the biased lensing power spectrum estimated from data. 
 
 Systematic contamination of the estimator by  the SZ signal, IR and radio sources was estimated in \citetalias{das/etal:2011a} using the simulations from  \citet{sehgal/etal:2010a}.  The authors found that, with the ACT lensing pipeline as used in this work, the contamination is smaller than the signal by two orders of magnitude and can thus be neglected. This result appears well-motivated for two reasons, which also apply to our current analysis: first, in the analysis we only use the signal-dominated scales below $\ell=2300$, at which SZ, IR and radio power are  subdominant; second, by using the data to estimate the bias, our estimator automatically subtracts the Gaussian part of the contamination, so that only a very small non-Gaussian residual remains. The previous ACT contamination estimates are not strictly applicable to the new lensing estimate, because the filters used here contain somewhat lower noise, and thus admit slightly more signal at higher $\ell$s; however, estimates by the SPT collaboration (van Engelen et al. 2012) with similar noise levels and filters also find negligible contamination. The contamination levels in our analysis are thus also expected to be negligible.

  The measured CMB lensing power spectrum, detected at  $\lensSigma\sigma$, is shown in Fig.~\ref{fig:cmbLens},  along with a theory curve showing the convergence power spectrum for a fiducial $\Lambda$CDM model defined by the parameter set $(\Omega_b, \Omega_m, \Omega_\Lambda, h, n_s,\sigma_8)$ =  (0.044, 0.264, 0.736, 0.71, 0.96, 0.80). Constraining the conventional lensing parameter $A_L$ that rescales the fiducial convergence power spectrum ($\cl^{\kappa\kappa} \rightarrow A_L \cl^{\kappa\kappa}$) we obtain $A_L = 1.06 \pm 0.23$. The data are thus a good fit to the $\Lambda$CDM prediction for the amplitude of CMB lensing. As in \citetalias{das/etal:2011a}, we find the spectrum to have Gaussian errors, uncorrelated between bins. 
  For some parameter runs, the lensing power spectrum information  is added to the CMB power spectrum information \citep{sievers/etal:prep}. 
%\subsection{Fields used}
%\subsection{Treatment of point sources --- inpainting/template sub?}
%\section{CMB convergence power spectrum results}

\section{Conclusions}
\label{sec:conclusions}

We have derived the power spectrum of microwave sky maps at 148 GHz
and 218 GHz produced by the Atacama Cosmology Telescope experiment,
such as those displayed in Fig. 5. The power spectra cover a range of
angular scales spanning nearly a factor of 20, ranging from around
0.35 degrees ($\ell=590$) to a little over one arcminute
($\ell=8900$). The maps are high quality, and in principle extracting
the power spectrum is a simple matter. A host of practical
considerations, along with the precision supported by the data, make
estimation of the power spectrum challenging. This paper summarizes
algorithms and techniques for handling the particular shapes of our
maps, point source contamination, the steepness of the power spectrum,
significant features due to galactic dust emission, spatially varying
noise levels, and calibration. In addition to the resulting power
spectra, we also display numerous null tests on the data. These tests,
along with results from simulated maps, make a strong case that any
systematic errors in our power spectra are below the level of
statistical error.

The ACT power spectra are consistent with those measured by the South
Pole Telescope collaboration, as are the underlying maps in a region
of overlapping sky coverage. Given how small the signals are and
how many sources of error must be tamed to measure them, consistent
results represent a substantial experimental achievement.

The temperature power spectrum measurements displayed in Figure~12
represent the culmination of a two-decade quest, since the first
large-angle power measurements were made by the COBE satellite \citep{smoot/etal:1992}.
 It was soon realized that for inflationary cosmological
models, the substantial structure in the microwave background
temperature angular power spectrum due to coherent acoustic
oscillations in the early universe would allow precise constraints on
the basic properties of the universe \citep{jungman/etal:1996}. A series
of innovative and increasingly sensitive experiments then gradually
unveiled the power spectrum. With the definitive measurements down to
quarter-degree scales by the WMAP satellite \citep{bennett/etal:prep} and
the precise arcminute-scale measurements by ACT (this work) and SPT
\citep{story/etal:prep} along with the  results  anticipated from the Planck satellite,
this particular route to cosmological insight is approaching a highly refined state.

A new frontier in microwave background experiments will likely be
detailed lensing maps from high-resolution polarization measurements
\citep{niemack/etal:2010, austermann/etal:2012}, which have the prospect
of constraining dark energy and modified gravity (e.g., \citealt{das/linder:2012}). 
The lensing power spectrum measurements presented here and by
SPT \citep{van/etal:2012} are the first steps along this new
path.

%\newpage
\acknowledgements 
This work was supported by the U.S. National Science Foundation through awards AST-0408698 and AST-0965625 for the ACT project, as well as awards PHY-0855887 and PHY-1214379. Funding was also provided by Princeton University, the University of Pennsylvania, and a Canada Foundation for Innovation (CFI) award to UBC. ACT operates in the Parque Astron\'omico Atacama in northern Chile under the auspices of the Comisi\'on Nacional de Investigaci\'on Cient\'ifica y Tecnol\'ogica de Chile (CONICYT). Computations were performed on the GPC supercomputer at the SciNet HPC Consortium. SciNet is funded by the CFI under the auspices of Compute Canada, the Government of Ontario, the Ontario Research Fund -- Research Excellence; and the University of Toronto.
SD acknowledges support from the David Schramm Fellowship at Argonne National Laboratory and the Berkeley Center for Cosmological Physics fellowship.
RD acknowledges support from FONDECYT and BASAL grants. 
\appendix
\section{Analytic errorbars}
\label{apdx:errorbars}
Here we derive an analytic expression for the expected error bars on the cross-frequency mulitseason cross-power spectrum. We denote the frequencies 
with uppercase $A$, $B$, $C$, $D$,  the seasons with  $\alpha$, $\beta$, $\gamma$, $\tau$, and the sub-season data splits with  $i$, $j$, $k$,$l$. 
Following \citetalias{das/etal:2011}, the covariance the of cross-power spectrum is defined as:
\begin{align}
 \Theta^{(\alpha A\times \beta B);( \gamma C\times \tau D)}_{bb}  \equiv \langle (C_{b,\alpha \beta} ^{(A \times B)}-  \langle C_{b,\alpha \beta} ^{(A \times B)} \rangle) (C_{b,\gamma \tau} ^{(C \times D)}-  \langle C_{b,\gamma \tau} ^{(C \times D)} \rangle) \rangle \nonumber ,\\
\end{align}
which expands as 
\begin{align}
\Theta^{ (\alpha A\times \beta B);( \gamma C\times \tau D)}_{bb}  = \frac{1}{N} \frac{1}{\nu_b^2} \sum_{i, j, k, l}^{n_{d}}  \sum_{\vec{\ell} \in b}   \sum_{ \vec{\ell'} \in b} \left( \left[\ave{T^{*iA}_{\vec{\ell}, \alpha} T^{jB}_{\vec{\ell}, \beta} T^{*kC}_{\vec{\ell}',\gamma} T^{lD}_{\vec{\ell}',\tau }}\right] - \ave{C_{b,\alpha \beta} ^{(iA \times jB)} } \ave{C_{b,\gamma \tau} ^{(kC \times lD)} } \right) \nonumberÊ\\ \times (1-\delta_{ij} \delta_{\alpha \beta})(1-\delta_{kl}  \delta_{\gamma \tau}). 
\end{align}
The Kronecker deltas remove the auto power spectra, and any same-split, same-season cross-frequency spectra. The general normalization is
\ba
N&=& \sum_{i, j, k, l}^{n_{d}}  (1-\delta_{ij}  \delta_{\alpha \beta})(1-\delta_{kl} \delta_{\gamma \tau})  \nonumber \\
&=&n_{d}^{4}-n_{d}^{3}( \delta_{\alpha \beta}+\delta_{\gamma \tau})+  n_{d}^{2}( \delta_{\alpha \beta}\delta_{\gamma \tau}). 
\ea
Applying Wick's Theorem, we have 
\begin{align}
\Theta^{(\alpha A\times \beta B);( \gamma C\times \tau D)}_{bb} = \frac{1}{\nu_b}  \frac{1}{N} \sum_{i, j, k, l}^{n_{d}} \left[\ave{C_{b, \alpha \gamma}^{iA\times kC}} \ave{C_{b, \beta \tau}^{jB\times lD}}  +  \ave{C_{b, \alpha \tau}^{iA\times lD}} \ave{C_{b, \beta \gamma}^{jB\times kC}}  \right]  \nonumberÊ\\ \times (1-\delta_{ij}  \delta_{\alpha \beta})(1-\delta_{kl}  \delta_{\gamma \tau}) ,
\end{align}
where 
\begin{align}
\ave{C_{b, \beta \gamma}^{jB\times kC}} = C_{b}+  \delta_{jk}\delta_{BC}\delta_{\beta \gamma} N_{b}^{\beta\beta,BB}.
\end{align}
Therefore, $\Theta^{(\alpha A\times \beta B);( \gamma C\times \tau D)}_{bb}$ expands to
\newcommand{\parenthnewln}{\right.\\&\left.\quad{}}
\begin{align}\begin{split} 
&\Theta^{1; (\alpha A\times \beta B);( \gamma C\times \tau D)}_{bb}  = 2 \frac{C_{b}^{2} }{\nu_{b}}  +  \frac{1}{N}\frac{C_{b}}{ \nu_{b}} \sum_{i, j, k, l}^{n_{d}} \left[( \delta_{ik}\delta_{AC}\delta_{\alpha \gamma} N_{b}^{\alpha \alpha,AA} + \delta_{jl}\delta_{BD}\delta_{\beta \tau} N_{b}^{\beta\beta,BB}) \parenthnewln +   ( \delta_{il}\delta_{AD}\delta_{\alpha \tau} N_{b}^{\alpha\alpha,AA} + \delta_{jk}\delta_{BC}\delta_{\beta \gamma} N_{b}^{\beta\beta,BB}) \right] \times (1-\delta_{ij}  \delta_{\alpha \beta})(1-\delta_{kl} \delta_{\gamma \tau}) \\ & + \frac{1}{N} \frac{1}{ \nu_{b}} \sum_{i, j, k, l}^{n_{d}} N_{b}^{\alpha\alpha,AA} N_{b}^{\beta\beta,BB} ( \delta_{ik}\delta_{AC}\delta_{\alpha \gamma} \delta_{jl}\delta_{BD}\delta_{\beta \tau} + \delta_{il}\delta_{AD}\delta_{\alpha \tau}  \delta_{jk}\delta_{BC}\delta_{\beta \gamma} )  \times (1-\delta_{ij} \delta_{\alpha \beta})(1-\delta_{kl} \delta_{\gamma \tau}),
\end{split}\end{align}
which after some algebra reduces to
\begin{align}\begin{split} 
&\Theta^{(\alpha A\times \beta B);( \gamma C\times \tau D)}_{bb}  =\frac{1}{\nu_{b}} \left( 2 C_{b}^{2} + \frac{ C_{b}N_{b}^{\alpha\alpha,AA} }{n_{d}}(\delta_{AC}\delta_{\alpha \gamma}+ \delta_{AD}\delta_{\alpha \tau})+\frac{C_{b} N_{b}^{\beta\beta,BB}}{n_{d} } ( \delta_{BC}\delta_{\beta \gamma}+\delta_{BD}\delta_{\beta \tau})  \parenthnewln+ N_{b}^{\alpha\alpha,AA} N_{b}^{\beta\beta,BB} (\delta_{AD}\delta_{\alpha \tau} \delta_{BC}\delta_{\beta \gamma}+\delta_{AC}\delta_{\alpha \gamma} \delta_{BD}\delta_{\beta \tau} ) \frac{n_{d}^{2} -n_{d}(\delta_{\alpha \beta}+ \delta_{\gamma \tau}) +n_{d}(\delta_{\alpha \beta}\delta_{\gamma \tau})}{n_{d}^{4}-n_{d}^{3}( \delta_{\alpha \beta}+\delta_{\gamma \tau})+  n_{d}^{2}( \delta_{\alpha \beta}\delta_{\gamma \tau})} \right)
\end{split}\end{align}

Therefore, with $A \ne B \ne C$ and $\alpha  \ne \beta$ we have,
\begin{align}
\Theta^{Ê(\alpha A\times \alpha A); (\alpha A\times \alpha A)}_{b} &= \frac{1}{ \nu_{b}} \left[ 2 C_{b}^{2} +4 \frac{C_{b}}{n_{d}}N_{b}^{\alpha \alpha, AA}+2 \frac{(N_{b}^{\alpha \alpha, AA})^{2}}{n_{d}(n_{d}-1)} \right] ,\\
\Theta^{Ê(\alpha A\times \alpha B);Ê(\alpha A\times \alpha B)}_{b}& = \frac{1}{ \nu_{b}} \left[ 2 C_{b}^{2}+ \frac{C_{b}}{n_{d}}( N_{b}^{\alpha \alpha, AA}+N_{b}^{\alpha \alpha, BB})+\frac{N_{b}^{\alpha \alpha ,AA}N_{b}^{\alpha \alpha, BB}}{n_{d}(n_{d}-1)}\right], \\
\Theta^{Ê(\alpha A\times \alpha A);Ê(\alpha A\times \alpha B)}_{b}&=  \frac{1}{\nu_b} \left[ 2C_{b}^{2} + 2\frac{C_{b}N_{b}^{\alpha \alpha, AA}}{n_{d}} \right] ,\\
\Theta^{Ê(\alpha A\times \alpha A);Ê(\alpha B\times \alpha B)}_{b}&=  \frac{1}{\nu_b} \left[ 2C_{b}^{2} \right], \\
\Theta^{Ê(\alpha A\times \alpha B);Ê(\alpha A\times \alpha  C)}_{b} &=\frac{1}{\nu_b}  \left[ 2C_{b}^{2} + \frac{C_{b}N_{b}^{\alpha \alpha, AA}}{n_{d}} \right] ,\\
\Theta^{Ê(\alpha A\times \beta A);Ê(\alpha A\times \beta B)}_{b} &=\frac{1}{\nu_b} \left[ 2C_{b}^{2} + \frac{C_{b}N_{b}^{\alpha \alpha, AA}}{n_{d}} \right] ,\\
\Theta^{Ê(\alpha A\times \beta A);Ê(\alpha B\times \beta B)}_{b}  &=\frac{1}{\nu_b} \left[ 2C_{b}^{2} \right] ,\\
\Theta^{Ê(\alpha A\times \beta B); (\alpha A\times \beta C)}_{b}  &=\frac{1}{\nu_b}  \left[ 2C_{b}^{2} + \frac{C_{b}N_{b}^{\alpha \alpha, AA}}{n_{d}} \right].
\end{align}

\section{Including beam covariance in the estimate}

We can also include the effect of an uncertainty on the beam when combining data between different season.  We consider the covariance of the power spectrum for two season pairs: $i\times j$s and $k \times l$, with beam window functions  $w^{i\times j}_{b}=B_{b}^{i}B_{b}^{j}$ and $w^{k \times l}_{b'}=B_{b'}^{k}B_{b'}^{l}$, respectively. The measured windows function is given by $w^{obs}_{b}=w_{b}+\delta w_{b}$.  This error propagates 
to the power spectrum covariance as
\ba
\langle  \frac{ w^{i\times j}_{b} C^{i\times j}_{b}}{w^{i\times j}_{b}+\delta w^{i\times j}_{b}}  \frac{ w^{k\times l}_{b'} C^{k\times l}_{b'}}{w^{k\times l}_{b'}+\delta w^{k\times l}_{b'}}    \rangle \rightarrow  \frac{ C^{i\times j}_{b}}{w^{i\times j}_{b}} \frac{ C^{k\times l}_{b'}}{w^{k\times l}_{b'}} \langle \delta w^{i\times j}_{b}  \delta w^{k\times l}_{b'}  \rangle .
\ea
The error on the window function is related to the error of the beam by
\ba
\delta w^{i\times j}_{b}= \delta B_{b}^{i}B_{b}^{j} +B_{b}^{i} \delta B_{b}^{j}.
\ea

\bibliography{act, wmap_supp}
\end{document}